\documentclass[aps,prd,preprint,eqsecnum,nofootinbib,groupedaddress,floatfix]{revtex4}

\usepackage{amsmath}
\usepackage{graphicx}
\usepackage{multirow}
\usepackage{placeins}
\usepackage{wasysym}
\usepackage{hyperref}
\hypersetup{
    colorlinks=true,
    allcolors={blue!60!black}
}
\usepackage{array}
\usepackage{todonotes}

\newcommand{\BlackHat}{{\sc BlackHat}}

\newcommand{\SHERPA}{{\sc Sherpa}}

\newcommand{\COMIX}{{\sc Comix}}
\newcommand{\SISCone}{{\sc SISCone}}
\newcommand{\ntuple}{{$n$-tuple}}

\newcommand{\Ntuples}{{$N$-tuples}}
\newcommand{\mc}[1]{\mathcal{#1}}

\newif\ifdraft
\draftfalse
\newif\ifpreprint
\preprinttrue

\def\fig#1{fig.~{\ref{#1}}}

\def\Wmj{$W^-\,\!+\,1$}
\def\Wmjj{$W^-\,\!+\,2$}

\def\Wjnp1{$W\,\!+\,(n\!+\!1)$}
\def\Wmjn{$W^-\,\!+\,n$}
\def\Wmjnm{$W^-\,\!+\,(n\!-\!1)$}
\def\Wpjn{$W^+\,\!+\,n$}

\def\Zjn{$Z\,\!+\,n$}

\def\jet{{\rm jet}}

\def\pT{p_{\rm T}}

\def\HTpartonic{{\hat H}_{\rm T}}
\def\HTpartonicp{{\hat H}_{\rm T}'}
\def\HTpartonicpp{{\hat S}_{\rm T}}

\def\MINLO{{\tt MiNLO}}
\def\MINLOp{{\tt MiNLO$^\prime$}}

\def\MILOp{{\tt MiLO$^\prime$}}
\def\MILNLO{{\tt Mi(N)LO}}
\def\MILNLOp{{\tt Mi(N)LO$^\prime$}}

\newbox\charbox
\newbox\slabox
\def\s#1{{      
	\setbox\charbox=\hbox{$#1$}
	\setbox\slabox=\hbox{$/$}
	\dimen\charbox=\ht\slabox
	\advance\dimen\charbox by -\dp\slabox
	\advance\dimen\charbox by -\ht\charbox
	\advance\dimen\charbox by \dp\charbox
	\divide\dimen\charbox by 2
	\raise-\dimen\charbox\hbox to \wd\charbox{\hss/\hss}
	\llap{$#1$}
}}

\begin{document}
\title{
	\ifpreprint
	\hbox{\rm\small
		FR-PHENO-2017-008$\null\hskip 3.0cm \null$
		SLAC--PUB--16945$\null\hskip 4.2cm \null$
	IPPP/17/99\break}
	\hbox{$\null$\break}
	\fi
	Weak Vector Boson Production with Many Jets at the LHC $\sqrt{s}= 13$ TeV}

	\author{F. R.~Anger${}^{a}$, F.~Febres Cordero${}^{a}$, S.~H{\"o}che${}^b$ and D.~Ma\^{\i}tre${}^c$
	\\
	$\null$
	\\
	${}^a${Physikalisches Institut, Albert-Ludwigs-Universit\"at Freiburg, D--79104 Freiburg, Germany}\\
	${}^b$SLAC National Accelerator Laboratory, Menlo Park, CA, 94025, USA\\
	${}^c$Department of Physics, University of Durham, Durham DH1 3LE, UK\\
}

\begin{abstract}
  Signatures with an electroweak vector boson and many jets play a crucial role
at the Large Hadron Collider, both in the measurement of Standard-Model
parameters and in searches for new physics. Precise predictions for these
multi-scale processes are therefore indispensable. We present next-to-leading
order QCD predictions for $W^\pm/Z$+jets at $\sqrt{s}=13$ TeV, including up to five/four
jets in the final state.  All production channels are included and leptonic
decays of the vector bosons are considered at the amplitude level. We assess
theoretical uncertainties arising from renormalization- and factorization-scale
dependence by considering fixed-order dynamical scales based on the $H_{\rm T}$
variable as well as on the \MINLO{} procedure. We also explore
uncertainties associated to different choices of parton-distribution functions. We provide
event samples that can be explored through publicly available \ntuple{} sets,
generated with \BlackHat{} in combination with \SHERPA{}.
\end{abstract}

\maketitle

\section{Introduction}
During the 7 and 8 TeV runs of the Large Hadron Collider (LHC), the ATLAS and CMS
experiments have scrutinized electroweak vector boson production in association
with multiple light jets in great detail~\cite{AtlasVjets7,CMSVjets8,CMSVjets7,ExptJetProductionRatioLHC}.
Recently, both collaborations presented first
measurements~\cite{AtlasVjets13,CMSVjets13} of this class of processes at
an energy of 13 TeV. These studies have
shown the extent to which theoretical predictions produced both by dedicated
calculations or by general Monte Carlo event generators can describe total rates
and differential distributions for signatures that involve leptons, missing
transverse energy and many light jets. Such characterizations are of key importance
given that searches for new physics carried out at hadron colliders
target similar final states.

Precise calculations for vector-boson production with a single jet have reached a new
era with the recent results at next-to-next-to-leading order (NNLO) QCD precision
for $V+1$ jet~\cite{VjetNNLO} as well as many other refinements, which for
example have been used to precisely describe backgrounds for dark matter
searches~\cite{Lindert:2017olm}. At next-to-leading-order (NLO) QCD, the state
of the art is nowadays calculations for processes with four and five
jets~\cite{W4j,W5j,Z4j} in the final state.  The matching to parton showers
has been carried out for up to three jets in the final state~\cite{MCatNLO},
and multi-jet merging has been studied with up to two jets~\cite{MEPSNLO}.
Next-to-leading order electroweak corrections have been computed for up to
three jets~\cite{WjetsNLOEW}, and combined with QCD merging~\cite{WjetsNLOEWmerge}.

In this article we extend the set of NLO QCD calculations by dedicated
predictions for the 13~TeV LHC for $W+n$-jet and $Z+m$-jet production with
$n\leq 5$ and \mbox{$m\leq 4$}.
We employ the \BlackHat{} library~\cite{BlackHatI} for computing the required
one-loop matrix elements. This library is based on on-shell and unitarity
techniques~(see for example~\cite{RecentOnShellReviews}) which allows the
extraction of loop amplitudes from simpler building blocks.  It has been
employed for many multi-light-jet studies at hadron
colliders~\cite{W3jPRL,W3jDistributions,BlackHatZ3jet,W4j,W5j,Z4j,BHother,Wratios},
and has recently been extended to applications for high-multiplicity processes
including heavy jets~\cite{Anger:2017glm}.
The calculation is performed with the help of the \SHERPA{} package~\cite{Sherpa},
used for integration over phase space as well as calculation of real corrections
employing the Catani-Seymour dipole subtraction~\cite{CS} as implemented in the
matrix-element generator \COMIX{}~\cite{Comix}. We have stored our results and
made them publicly available as Root~\cite{ROOT} \ntuple{} files. This file
format~\cite{BHNtuples} contains all information necessary to compute the NLO
fixed-order predictions, to change the running QCD coupling, renormalization and
factorization scales, as well as parton-distribution functions (PDFs). We employ
an extension~\cite{Greiner:2016awe} of the \ntuple{} file format which allows for
extended reweighting procedures.

Due to its large center-of-mass energy, the LHC explores scales
ranging from tens of GeV's to the multi-TeV regime.
This large hierarchy of scales makes leading-order (LO) QCD predictions unreliable, as they tend
to be very sensible to the unphysical renormalization and factorization scales,
and also because they can miss important initial-state partonic configurations.
Next-to-leading order QCD calculations are less affected by scale choices
and give a first reliable estimate of the theoretical uncertainty\footnote{
  A notable exception to this are processes with large accessible phase space
  and missing partonic channels at the next-to-leading order. This has been
  pointed out for example in~\cite{RubinSalamSapeta}.}.
We study in detail in this paper the theoretical uncertainties associated
to our predictions that are related to scale sensitivity and PDF dependence.
We explore in particular the scale dependence by using fixed-order dynamical
scales based on the total partonic transverse energy
as well as different variants of the \MINLO{} method~\cite{MINLO}. In addition,
we perform the conventional scale variation by constant factors around the central
scales. A similar study was carried out recently for on-shell $t\bar t$ production
in association with up to three light jets~\cite{ttjjj}, and good agreement was found
between NLO QCD results employing a similar set of dynamical scales. Our analysis
is the first to compare results obtained with fixed-order scales and with the
\MINLO{} method in processes with four or five light jets in the final state.

This paper is organized as follows. In section~\ref{sec_setup} we summarize our
calculational setup, showing all kinematical information employed both at the
level of producing \ntuple{} files and for the distributions studied in the rest
of the paper. We also show in this section our implementation of the dynamical
scales employed. In section~\ref{sec_vjets} we present our results for total and
differential cross sections and study their scale dependence and uncertainties
associated to PDFs. In section~\ref{sec_ratios} we show a series of observable
ratios that can help to reduce the theoretical uncertainties in multi-jet
environments. The paper ends with our conclusions in section~\ref{ConclusionSection}.


\section{Basic Setup}
\label{sec_setup}

We employ the \SHERPA{} package~\cite{Sherpa} to manage the overall
calculations. The required one-loop matrix elements are produced by the
\BlackHat{} library~\cite{BlackHatI}.
Born and real-emission contributions are computed by the matrix-element
generator \COMIX{}~\cite{Comix}, which also
provides the necessary Catani-Seymour subtraction terms~\cite{CS}. More details
of our computational setup can be found in Refs.~\cite{W5j} and~\cite{Z4j}. We
have included in all of our results all contributing subprocesses, confirming in
particular that 8-quark finite contributions to the real part in $W^\pm+5$-jet
production are negligible~\cite{W5j}.

We use the {\tt CT14} LO ({\tt CT14llo}) and NLO ({\tt CT14nlo})
PDFs~\cite{CT14} at the respective orders, including the corresponding
definition of the strong coupling $\alpha_s$. We also employ the corresponding
{\tt CT14nlo} error set to explore PDF uncertainties, and compare to predictions
generated with the PDF error sets of ABM~\cite{ABM}, MMHT~\cite{MMHT} and NNPDF 3.1~\cite{NNPDF}.
The lepton-pair invariant mass follows a relativistic Breit-Wigner distribution,
with $M_W=80.385$ GeV and $M_Z=91.1876$ GeV, and the widths are given by
$\Gamma_Z=2.4952$~GeV and $\Gamma_W=2.085$~GeV. We employ a diagonal CKM
matrix and use real values for the electroweak parameters.

All light quarks ($u$, $d$, $s$, $c$, $b$) are treated as massless. We do not
include contributions from real or virtual top quarks, and we expect this to have
a percent-level effect on cross-sections~\cite{W4j,Z4j,Campbell:2016tcu,Anger:2017glm}.  The
one-loop matrix elements for $V+4,5$-jets have been computed in the
leading-color approximation~\cite{Ita:2011ar} which we find to be
precise at the level of 2\% of the total cross section in lower jet-multiplicity calculations.  Our results are quoted for a single 
lepton (pair) flavor. We treated both leptons as massless, an approximation that can be 
applied to the electron or muon families.
Results presented are produced in fixed-order parton-level perturbation theory
and we do not apply any non-perturbative corrections to account for effects
associated to underlying event or hadronization.

In the following subsections we describe first the common setup to define the
fiducial regions in which total and differential cross sections are computed.
Second, we list the basic phase-space cuts applied at the level of producing the \ntuple{}
sets and finally the dynamical scales that we employ to study the scale sensitivity
of our results.

\subsection{Kinematical Setup}
\label{sec_kin}

In our study we consider the inclusive processes $pp\rightarrow V+n$ jets at the
LHC with center-of-mass energy $\sqrt{s}=13$ TeV, with $n\le 5$ and $n\le 4$ for
$V=W^\pm$ and $Z$, respectively. We define jets using the anti-$k_{\rm T}$
algorithm~\cite{antikT} with $R=0.4$ and impose the kinematical cuts:
\begin{eqnarray}
&& \pT^\jet > 30 \hbox{ GeV}\,, \hskip 1.5 cm 
|\eta^\jet|<3\,. 
\label{eq_jets}
\end{eqnarray}
We order the jets in $p_{\rm T}$ and label them according to their hardness. For
all charged leptons we require:
\begin{eqnarray}
&& \pT^l > 20 \hbox{ GeV}\,, \hskip 1.5 cm 
|\eta^l|<2.5\,.
\label{eq_clep}
\end{eqnarray}
For processes with $W^\pm$ bosons we define its transverse mass by $M_{\rm
T}^W=\sqrt{2E_{\rm T}^lE_{\rm T}^\nu(1-\cos(\Delta \phi_{l\nu}))}$. For these we
impose the additional cuts:
\begin{eqnarray}
&& \pT^\nu > 20 \hbox{ GeV}\,, \hskip 1.5 cm 
M_{\rm T}^W>20 \hbox{ GeV}\,.
\label{eq_W}
\end{eqnarray}
Finally for processes with a $Z$ boson we impose the following constraint on the
invariant mass of its decay products:
\begin{equation}
66 \hbox{ GeV}<M_{l^+l^-}<116 \hbox{ GeV}\,.
\label{eq_Z}
\end{equation}

\subsection{Kinematical Setup for Public \Ntuples{}}
\label{sec_kin_ntuples}

We have saved intermediate results in publicly available Root-format~\cite{ROOT}
\ntuple{} files~\cite{BHNtuples} in order to facilitate new studies of
infrared-safe observables  with different cuts, different scale choices, jet
algorithms or PDF sets. Our study uses an extension of the original file format
which facilitates reweighting in the \MINLO{} procedure and enables extended
reweighting procedures~\cite{Greiner:2016awe}. Due to the generation cuts
applied to events stored in the \ntuple{} files, future studies should operate
either with identical cuts or with tighter cuts.

We use the {\sc FastJet} package~\cite{FastJet} to define jets according to
the algorithms anti-$k_{\rm T}$, $k_{\rm T}$ and
\SISCone{}~\cite{antikT,KT,SISCONE} with jet parameter $R=0.4,0.5,0.6$ and
$0.7$. We set the recombination $f$-parameter for the \SISCone{} algorithm to
$0.75$. We impose $p_{\rm T}^{\rm jet}>30$ GeV on all jets and
do not constrain their pseudorapidity.

For charged leptons we impose $p_{\rm T}^l>20$ GeV and $|\eta^l|<3$, while for
neutrinos we require $p_{\rm T}^\nu>10$ GeV. For processes with a $Z$
boson, in which $\gamma^*$ contributions appear, we also constrain the
associated invariant-lepton-pair mass by $60\ {\rm GeV}<M_{l^+l^-}<120$ GeV.

\subsection{Dynamical scale choices}
\label{sec_scales}

We explore the renormalization and factorization scale dependence of the
cross sections using a conventional variation of the central scale by factors
$(1/2,1/\sqrt{2},1,\sqrt{2},2)$, keeping factorization and renormalization
scales equal. In addition, we explore the sensitivity of the calculation to
the functional form of the scale using on one hand conventional
fixed-order scales based on the $H_{\rm T}$ variable and on the other hand the
so-called \MINLO{} procedure~\cite{MINLO}.
In this subsection, we first define all the fixed-order scales employed,
and second we present our variant of the \MINLO{} procedure, which we
will label \MINLOp{}. Finally we summarize the nomenclature used in
this article to label all dynamical scales considered.

\subsubsection{Fixed-order scales}
\label{sec_fo_scales}
We define the total partonic transverse energy variable $\HTpartonicp$ according
to:
\begin{equation}
\HTpartonicp=\sum_j p_{\rm T}^j+E_{\rm T}^V\ ,
\label{eq_HTp}
\end{equation}
where the sum runs over all final-state partons and $E_{\rm T}^V\equiv
\sqrt{M_V^2+(p_{\rm T}^V)^2}$ is the transverse energy of the vector boson ($V$
either $W$ or $Z$). The scale $\mu_0=\HTpartonicp/2$ has proven to be a sensitive
choice as it tends to reduce the shape changes and global size of quantum corrections
when going from leading to next-to-leading order (see for example~\cite{W5j,Z4j,Wratios}).
In general, NLO corrections are less sensitive to the choices of scale, as long
as the scale reflects the hardness of the Born process~\cite{W3jDistributions}.

We introduce an additional scale, designed to match the invariant mass of the
lepton pair in kinematic configurations with very small hadronic transverse energy
and the transverse momentum of the hardest QCD jet in processes of di-jet type.
This scale is denoted as
\begin{equation}
\HTpartonicpp=\frac{1}{2}\sum_j p_{\rm T}^j+E_{\rm T}^V\ .
\label{eq_HTpp}
\end{equation}
In section~\ref{sec_scale_notation} we give the standard notation used
throughout this article to refer to the different scales employed.

\subsubsection{\MINLOp{}}
\label{sec_minlo_gen}
The second type of dynamical scale considered in this study, schematically denoted as \MINLOp{},
is based on the \MINLO{} reweighting procedure proposed in Ref.~\cite{MINLO}, which is inspired
by the NLL branching formalism in~\cite{KT}. It builds on an event-by-event identification
of the most likely branching history leading to the full $V+n-$parton final state using
a $k_T$-type clustering algorithm. No-branching probabilities in the form of NLL
Sudakov form factors are assigned to the intermediate ``partons'' in the branching tree,
to reflect the fact that no radiation above a resolution scale, given by lowest nodal
$k_T$ value in the clustering, should occur. The strong coupling associated to each node
in the branching tree is evaluated at the respective transverse momentum, following~\cite{Amati:1980ch}.
This method can also be interpreted as a generalization of the CKKW procedure~\cite{CKKW}
to NLO QCD calculations, but without the possibility to further develop the intra-jet real
radiation pattern through resolved emissions. In particular, the \MINLO{} method accounts
for the resummation of large logarithmic corrections associated to very disparate scales
in high-energy collisions using the known universal factorization properties of the
cross section in the collinear limit.

In order to reflect the nature of QCD interactions, only $1\rightarrow 2$ branchings
consistent with elementary interaction vertices are allowed in our
\MINLOp{} procedure. In addition, we require the branching history to be ordered,
which implies that we terminate the clustering as soon as an inverted scale hierarchy
is encountered. In this case, by default we set the scale of the remaining $V+m$-parton (with $m<n$) ``core'' interaction,
$\mu_{\rm core}$, to $\HTpartonicp/2$ (or when explicitly stated, to $\HTpartonicpp$).
This biases the scale choice for events with many hard scales towards $\HTpartonicp/2$
($\HTpartonicpp$), an effect that will be further discussed in Sec.~\ref{sec_vjets}.
Note in particular that at very high energies there may be configurations where no clustering
can be performed at all, for example a $V+2$ jet event with
$p_{T,j1}\approx p_{T,j2}\gg m_{T,W}$ or a $V+5$ jet event where
$p_{T,j1}\approx p_{T,j2}\approx\ldots\approx p_{T,j5}$
and $y_{j1}\ll y_{j2}\ll\ldots\ll y_{j5}$. The large logarithms associated with such
configurations cannot be resummed in QCD collinear factorization and are therefore not
amenable to a treatment in the coherent branching formalism on which the \MINLO{} method is based.
This is a considerable source of uncertainty because of the large available
phase space at the LHC.

In general, the clustering procedure for a leading-order process of
$\mathcal{O}(\alpha_s^N)$ will yield a branching history with $M\le N$
ordered nodal scales $q_1\ldots q_M$ and a core interaction of
$\mathcal{O}(\alpha_s^{N-M})$ with scale $\mu_{\rm core}>q_M$.
We then set the global renormalization scale $\mu_R$ to the geometric mean
$\mu_R^{N}=\mu_{\rm core}^{N-M}\prod_{i=1}^{M}q_i$.

Both intermediate lines (connecting branching nodes $i$ and $j$) and external
lines are dressed with Sudakov form factors to reflect the no-branching probability.
External lines connected to the $i$-th branching are multiplied by a
factor $\Delta_a (q_{min},q_i)$, where the lowest branching scale $q_1=q_{min}$ is
identified as the resolution scale. Intermediate lines connecting nodes $j<i$
are dressed by factors $\Delta_a (q_{min},q_i)/ \Delta_a (q_{min},q_j)$.
Internal lines connected to the primary process are assigned form factors
between their respective scales and $\mu_{\rm core}$. The factorization scale
$\mu_F$ used in the evaluation of the PDFs is set to the lowest scale, $\mu_F=q_1$.

In our \MINLOp{} procedure, we use a physical definition of the Sudakov form factors, which is given by
\begin{equation}\label{eq:def_nll_sudakov}
  \begin{split}
    &\Delta_a(Q_0,Q)=\exp\left\{-\int_{Q_0}^Q\frac{{\rm d} q}{q}
    \frac{\alpha_s(q)}{\pi}\sum_{b=q,g}\int_0^{1-q/Q}{\rm d}z
    \left(z\,P_{ab}(z)+\delta_{ab}\frac{\alpha_s(q)}{2\pi}\frac{2C_a}{1-z}K\right)\right\}\;,
  \end{split}
\end{equation}
where~\cite{Catani:1990rr}
\begin{equation}
  K=\left(\frac{67}{18}-\frac{\pi^2}{6}\right)C_A-\frac{10}{9}T_R\,n_f\;,
\end{equation}
and $a=g,q$ corresponds to massless gluons and quarks, respectively.
Eq.~\eqref{eq:def_nll_sudakov} does not exceed unity and can therefore
be interpreted as a no-branching probability between the scales
$Q_0$ and $Q$, while maintaining the correct limiting behavior for $Q_0\ll Q$.
The above definition is easily obtained from the known NLL expressions
of~\cite{Catani:1991hj} by employing the following symmetry of the LO DGLAP
splitting functions
\begin{equation}
  \begin{split}
    \sum_{b=q,g}\int_{0}^{1-\varepsilon} dz\, z\, P_{qb}(z)
    =&\int_{\varepsilon}^{1-\varepsilon} dz\, P_{qq}(z)+\mc{O}(\varepsilon)\;,\\
    \sum_{b=q,g}\int_{0}^{1-\varepsilon} dz\, z\, P_{gb}(z)
    =&\int_{\varepsilon}^{1-\varepsilon} dz\,\Big[\;\frac{1}{2}P_{gg}(z)+n_f\,P_{gq}(z)\;\Big]
    +\mc{O}(\varepsilon)\;.
  \end{split}
\end{equation}
Following standard practice, we include next-to-leading logarithms proportional
to the two-loop cusp anomalous dimension~\cite{Catani:1990rr}.

The generalization of the \MINLO{} method to NLO requires some modifications~\cite{MINLO}.
Virtual corrections and integrated IR-subtraction terms are treated identically
to the leading order case. Real-emission events have branching histories with
$M+1 \leq N+1$ ordered branchings, but they are treated as born-like $M$-parton events
for the purpose of scale definition. This is achieved by discarding the softest branching,
i.e. if the $M+1$ step branching history is given by $q_0<q_1<\dots<q_K$, we set the
resolution scale to $q_1$. Consequently, the softest emission at NLO (with scale $q_0$)
is neither dressed with Sudakov factors nor does it enter the definitions of $\mu_R$
and $\mu_F$. In order to retain NLO accuracy of the full calculation, the Born configuration
receives correction terms that are proportional to the first-order expansion of the Sudakov
factors, eq.~\eqref{eq:def_nll_sudakov}. Note that in this case the scale of the strong coupling
is set to $\mu_R$, and that the two-loop cusp term is neglected, as it contributes
at $\mathcal{O}(\alpha_s^2)$. The value of the additional strong coupling at NLO (the $N+1$-th power)
appearing in both real and virtual corrections is set to the average of all other values
of $\alpha_s$, i.e.\ $N\,\alpha_s^{(N+1)}=(N-M)\,\alpha_s(\mu_{\rm core})+\sum_{i=1}^M\alpha_s(q_i)$.
Conventional scale uncertainties associated to the \MINLO{} method are estimated
using variations of $\mu_R$ and $\mu_F$ by constant factors of two. The scale of
the strong coupling in the Sudakov form factors remains fixed at the integration variable, $q$,
while it is varied in all other parts of the calculation, including the \MINLO{} counterterms
used to subtract the $\mc{O}(\alpha_s)$ expansion of the Sudakov factors.
Factorization scale variations in the \MINLOp{} procedure have been
discussed extensively in~\cite{ttjjj}. We perform them in the same manner,
i.e.\ we set $q_1$ equal to $\mu_F$.

\subsubsection{Nomenclature for dynamical scales explored}
\label{sec_scale_notation}
Throughout this paper, we set renormalization and factorization scales equal
$\mu_R=\mu_F=\mu_0$. Our results labeled ``LO'' and ``NLO'' use the
central scale $\mu_0=\HTpartonicp/2$ by default, where $\HTpartonicp$ is defined
in eq.~\eqref{eq_HTp}. When necessary, to distinguish the usage of the fixed-order
scales defined in section~\ref{sec_fo_scales}, we write ``(N)LO
$\HTpartonicp/2$'' or ``(N)LO $\HTpartonicpp$'', where $\HTpartonicpp$ is
defined in eq.~\eqref{eq_HTpp}. 

In our \MINLOp{} procedure described in section~\ref{sec_minlo_gen} the default
core scale is $\mu_{\rm core}=\HTpartonicp/2$. When considering variations of
this choice we explicitly write ``\MILNLOp{}~$\HTpartonicp/2$'' or
``\MILNLOp{}~$\HTpartonicpp$''. 

We also compare to the original formulation of the \MINLO{}
method~\cite{MINLO} for processes with fewer than three jets in the
final state. Compared to our implementation, this variant
uses the Sudakov factors of~\cite{Catani:1991hj} and unordered clustering histories
are treated in a different manner.
Following the previous naming convention, we label those results as 
``\MILNLO{}~$\HTpartonicp/2$'' or ``\MILNLO{}~$\HTpartonicpp$'' depending on the
choice of core scale $\mu_{\rm core}$ employed.

\section{Results for $V$ + Jets production}
\label{sec_vjets}

\subsection{Total cross sections}
\label{totalxsw}

In tables~\ref{tab_Wpj_total_xs}, \ref{tab_Wmj_total_xs}
and~\ref{tab_Zj_total_xs} we present total cross sections for the production of
a weak vector boson $V$ in association with up to 5 jets for $V=W^+$ and $W^-$,
and with up to 4 jets for $V=Z$. Results with central scale
$\HTpartonicp/2$ and \MINLOp{} are included (see
section~\ref{sec_scale_notation} for the nomenclature that we use for the
different dynamical scales considered). We also show in
table~\ref{tab_jet_prod_total_xs} jet production
ratios~\cite{JetProductionRatio,Wratios} for all the vector boson, that is the
ratios of the total cross sections for the production for $V+n$ jets to the
production of $V+(n-1)$ jets.

\begin{table}[p]
\begin{center}
  \begin{ruledtabular}
    \begin{tabular}{ccccccc}
      jets  & $W^+$ LO & $W^+$ NLO & $W^+$ \MILOp{} & $W^+$ \MINLOp{} & \MILOp{} /
      LO &  \MINLOp{}/ NLO\\
      \noalign{\vskip 1mm}
      \colrule
      \noalign{\vskip 1mm}
      1 & $588.49(33)^{+23.77}_{-27.07}$ & $764.9(16)^{+37.8}_{-27.1}$ & $591.50(35)^{+20.70}_{-25.36}$ & $799.1(18)^{+49.9}_{-35.1}$ & $1.005(1)$ & $1.045(3)$ \\
      2 & $197.23(27)^{+44.64}_{-34.42}$ & $197.78(66)^{+1.80}_{-7.82}$ &$205.01(28)^{+46.95}_{-36.46}$ & $211.44(78)^{+9.50}_{-12.07}$ & $1.039(2)$ & $1.069(5)$ \\
      3 & $57.07(10)^{+22.82}_{-15.23}$ & $49.54(27)^{+0.00}_{-3.13}$ &
      $59.09(11)^{+26.02}_{-16.89}$ & $52.32(41)^{+0.19}_{-3.95}$ & $1.035(3)$
      & $1.056(10)$\\
      4 & $16.408(50)^{+9.344}_{-5.566}$ & $12.14(22)^{+0.00}_{-1.59}$ &
      $17.287(56)^{+11.768}_{-6.516}$ & $12.78(24)^{+0.00}_{-2.87}$ &$1.054(5)$ & $1.053(27)$\\
      5 & $4.579(45)^{+3.399}_{-1.829}$ & $3.06(14)^{+0.00}_{-0.72}$ &
      $4.908(56)^{+4.691}_{-2.233}$ & $3.21(14)^{+0.00}_{-0.52}$ &
      $1.072(16)$ & $1.049(66)$ 
    \end{tabular}
  \end{ruledtabular}
\end{center}
\caption{LO and NLO QCD results for inclusive $W^++1,2,3,4,5$-jet total cross sections
(in $pb$). Results are quoted for the two dynamical scales, $\HTpartonicp/2$ and
\MINLOp{}, as well as their ratio. Details of the calculation are given in section~\ref{sec_kin}.
The conventional scale dependence is determined by varying $\mu_R$ and $\mu_F$ by factors of $2$ and $\sqrt{2}$ up and down.
The number in parenthesis gives the statistical error from the numerical
integration. \label{tab_Wpj_total_xs} }
\end{table}

\begin{table}[p]
\begin{center}
  \begin{ruledtabular}
    \begin{tabular}{ccccccc}
      \noalign{\vskip 1mm}
      jets  & $W^-$ LO & $W^-$ NLO & $W^-$ \MILOp{} & $W^-$ \MINLOp{}
      &\MILOp{} / LO &\MINLOp{} / NLO \\
      \noalign{\vskip 2mm}
      \colrule
      \noalign{\vskip 2mm}
      1 & $446.68(22)^{+17.48}_{-20.29}$ & $582.0(11)^{+25.6}_{-18.6}$ &  $448.82(23)^{+15.07}_{-18.92}$ & $608.2(12)^{+34.6}_{-24.5}$ & $1.005(1)$ & $1.045(3)$ \\
      2 & $141.67(14)^{+31.89}_{-24.67}$ & $144.53(39)^{+1.03}_{-5.46}$ & $147.40(15)^{+33.40}_{-26.08}$ & $154.11(46)^{+5.91}_{-8.40}$& $1.040(1)$ & $1.066(4)$ \\
      3 & $39.029(55)^{+15.653}_{-10.448}$ & $34.34(16)^{+0.00}_{-2.10}$ & $40.889(61)^{+17.99}_{-11.70}$ & $36.20(22)^{+0.03}_{-2.64}$ & $1.048(2)$ & $1.054(8)$\\
      4 & $10.513(23)^{+6.035}_{-3.585}$ & $8.22(13)^{+0.00}_{-0.86}$ & $11.399(27)^{+7.797}_{-4.314}$ & $8.85(19)^{+0.00}_{-1.85}$ & $1.084(3)$ & $1.077(29)$ \\
      5 & $2.747(12)^{+2.059}_{-1.103}$ & $1.971(56)^{+0.004}_{-0.298}$ & $3.063(14)^{+2.949}_{-1.400}$ & $2.105(59)^{+0.000}_{-0.788}$ &$1.115(7)$ & $1.068(43)$\\
    \end{tabular}
  \end{ruledtabular}
\end{center}
\caption{As in table~\ref{tab_Wpj_total_xs} but for inclusive
  $W^-+1,2,3,4,5$-jet total cross sections.\label{tab_Wmj_total_xs} }
\end{table}

\begin{table}[p]
\begin{center}
  \begin{ruledtabular}
    {\footnotesize
    \begin{tabular}{ccccccc}
      \noalign{\vskip 1mm}
      jets  & $Z$ LO & $Z$ NLO  & $Z$ \MILOp{} & $Z$ \MINLOp{} & \MILOp{}
      / LO & \MINLOp{} / NLO  \\
      \noalign{\vskip 2mm}
      \colrule
      \noalign{\vskip 2mm}
      1 & $112.264(60)^{+4.121}_{-4.876}$ & $142.79(15)^{+5.12}_{-3.70}$ & $112.615(43)^{+3.390}_{-4.448}$ & $148.48(17)^{+7.21}_{-5.04}$ & $1.003(1)$ & $1.040(2)$\\
      2 & $36.140(38)^{+7.931}_{-6.178}$ & $36.811(65)^{+0.228}_{-1.339}$ & $36.780(28)^{+8.076}_{-6.382}$ & $38.962(68)^{+1.555}_{-2.147}$ & $1.018(1)$ & $1.058(3)$\\
      3 & $10.4844(76)^{+4.1227}_{-2.7702}$ & $9.175(44)^{+0.000}_{-0.578}$
      &  $11.1242(87)^{+4.847}_{-3.166}$ & $9.612(50)^{+0.000}_{-0.617}$ &
      $1.061(1)$ & $1.048(7)$\\
      4 & $2.9597(37)^{+1.6698}_{-0.9989}$ & $2.331(29)^{+0.000}_{-0.246}$ & $3.3050(43)^{+2.248}_{-1.247}$ & $2.439(37)^{+0.000}_{-0.668}$ & $1.117(2)$ & $1.046(21)$
    \end{tabular}}
  \end{ruledtabular}
\end{center}
\caption{As in table~\ref{tab_Wpj_total_xs} but for inclusive
  $Z+1,2,3,4$-jet total cross sections.\label{tab_Zj_total_xs} }
\end{table}

\begin{figure}[p]
\begin{center}
\includegraphics[clip,scale=0.45]{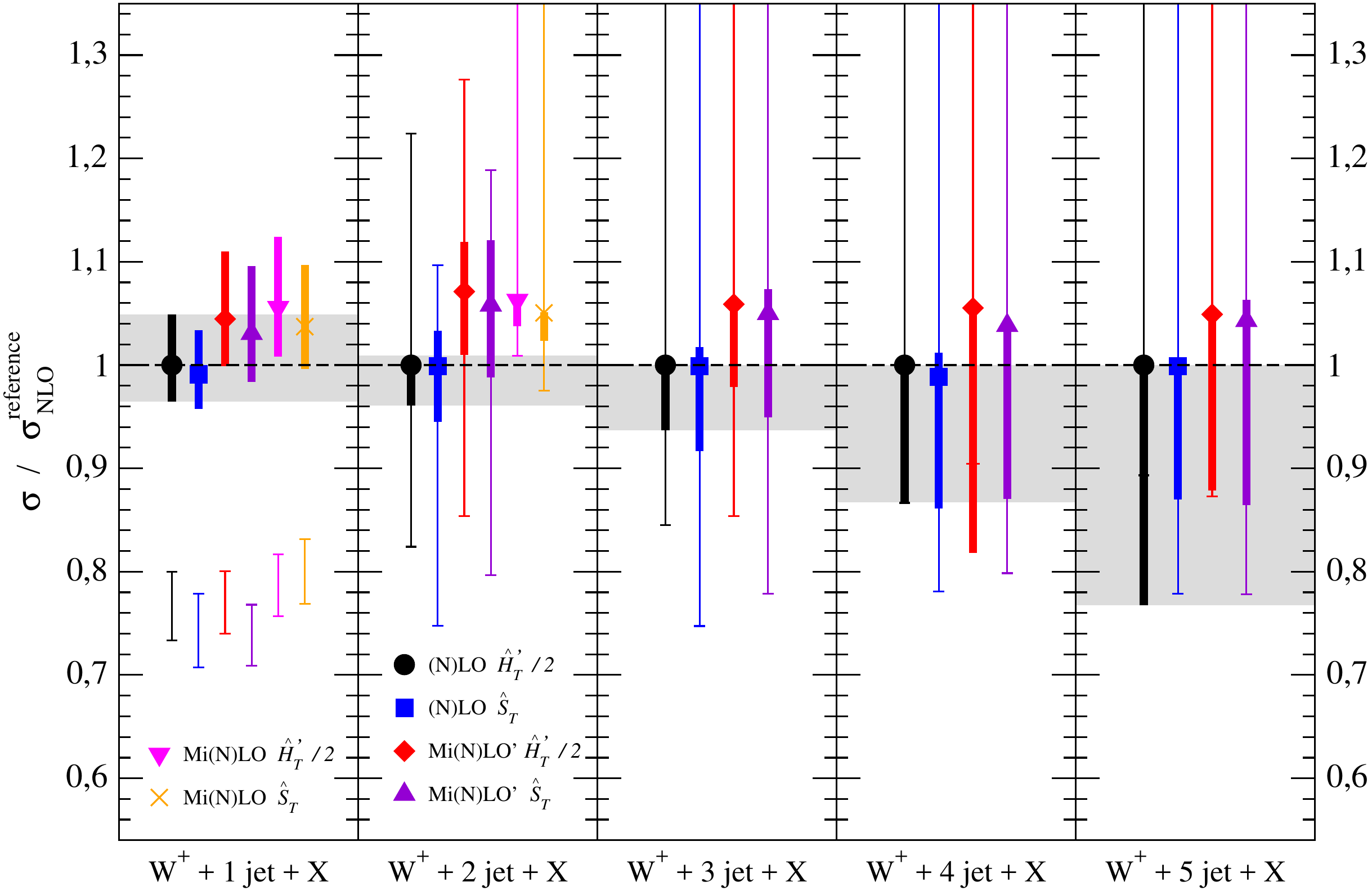}
\end{center}
\caption{The scale dependence of total cross sections for inclusive
$W^++1,2,3,4,5$-jet production considering all scales named in
section~\ref{sec_scale_notation}. Each scale is presented with a corresponding
symbol (and color) specified in the label of the plot. The thin lines represent
corresponding LO scale-dependence bands (in certain cases extending outside of
the range plotted) while the thick lines represent NLO bands. All
results have been normalized to the corresponding NLO $\HTpartonicp/2$ result at
each jet multiplicity.
The gray band show the NLO $\HTpartonicp/2$ scale-dependence band.}
\label{fig_Wp_all_scales}
\end{figure}

\begin{figure}[p]
\begin{center}
\includegraphics[clip,scale=0.45]{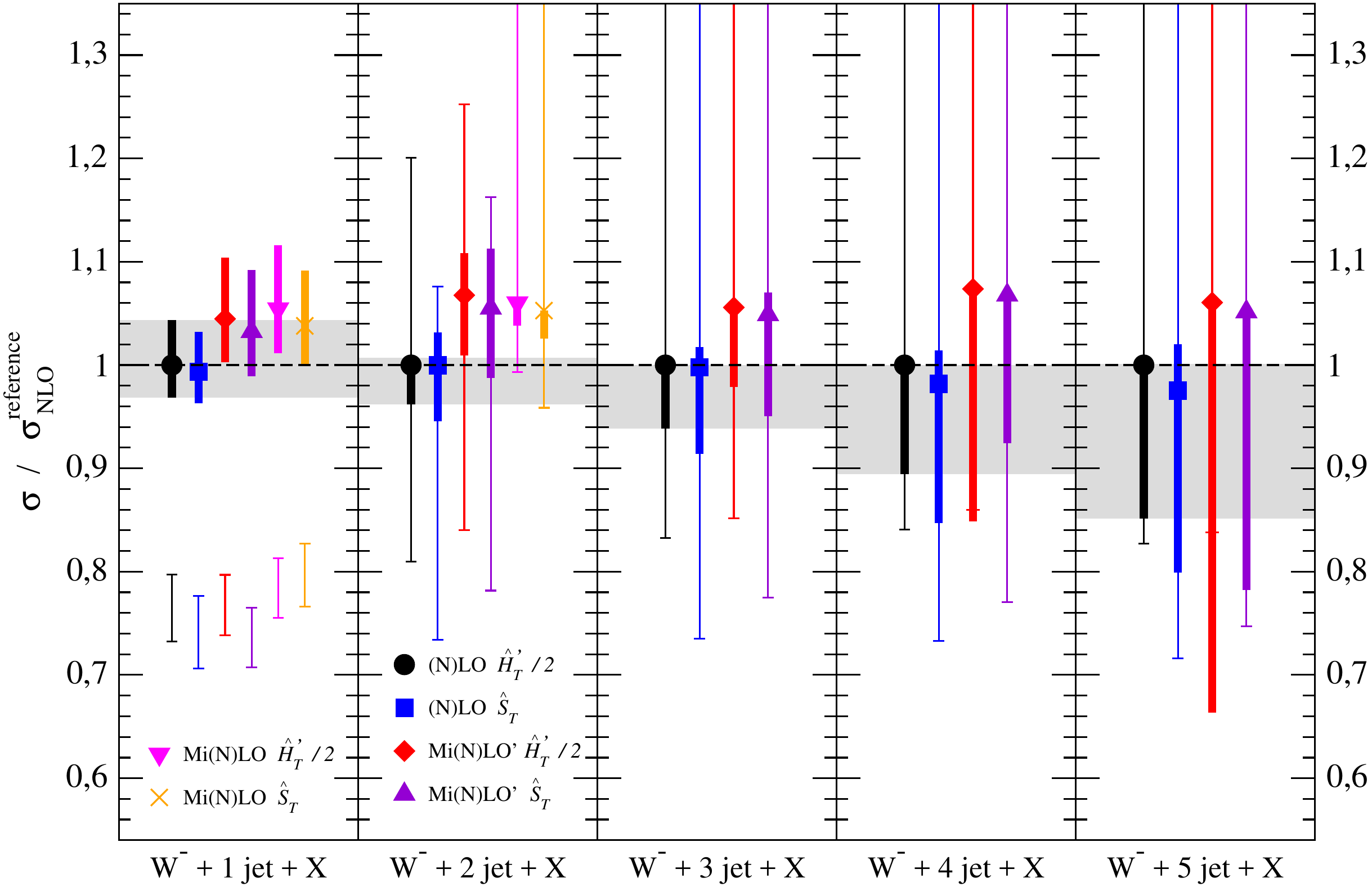}
\end{center}
\caption{As in \fig{fig_Wp_all_scales} but for inclusive $W^-$ production
in association with jets. See section~\ref{sec_scale_notation} for 
details of the dynamical scales considered.}
\label{fig_Wm_all_scales}
\end{figure}

\begin{figure}[t]
\begin{center}
\includegraphics[clip,scale=0.45]{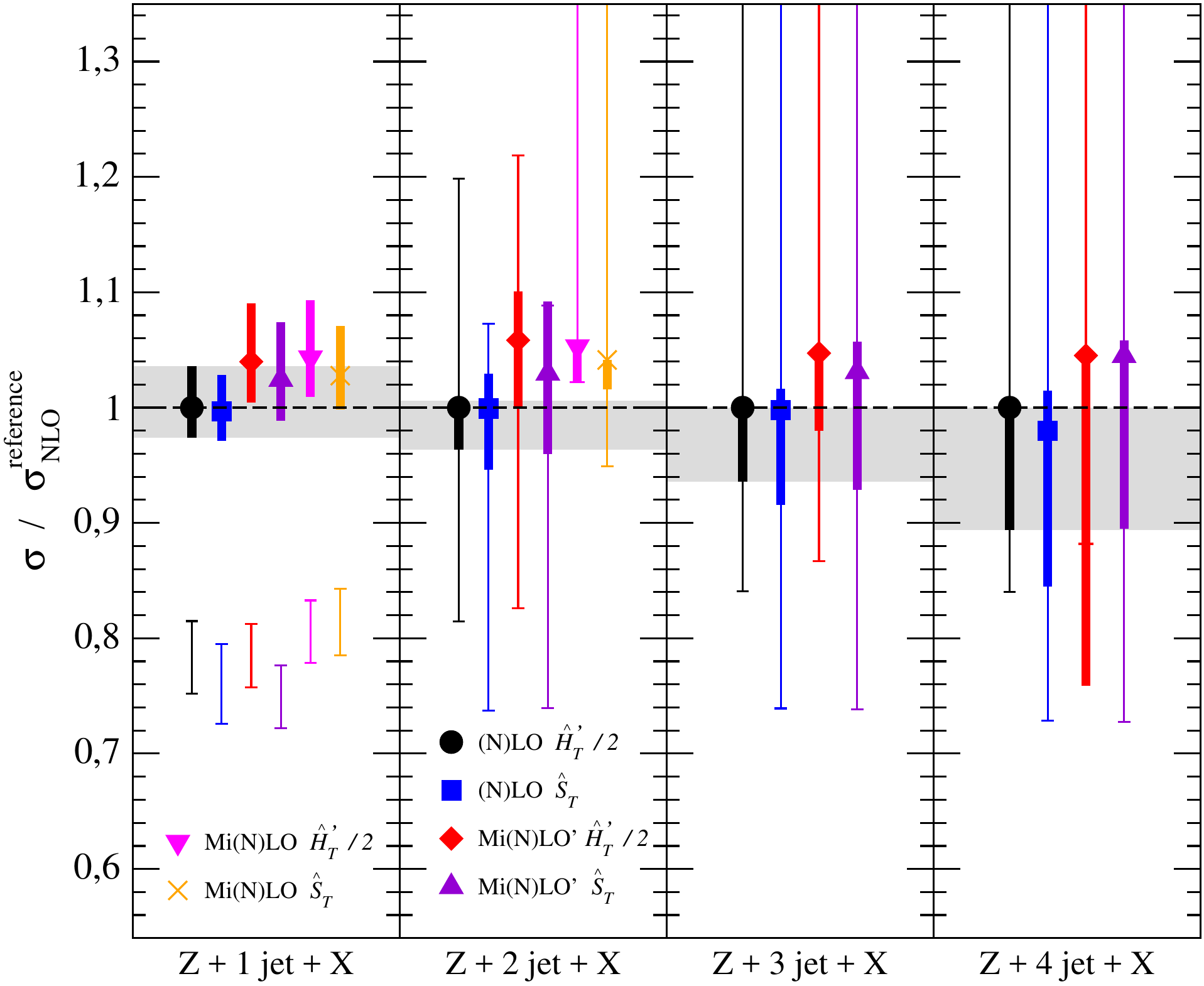}
\end{center}
\caption{As in \fig{fig_Wp_all_scales} but for inclusive $Z$ production
in association with jets. See section~\ref{sec_scale_notation} for 
details of the dynamical scales considered.}
\label{fig_Z_all_scales}
\end{figure}

\begin{table}[t]
  \begin{center}
    \begin{ruledtabular}
      \begin{tabular}{>{\centering\arraybackslash\hspace{0pt}}p{0.6cm}ccccccccc}
        \noalign{\vskip 1mm}
        \multicolumn{1}{c}{ } & \multicolumn{3}{c}{$W^+ n/(n-1)$} &
        \multicolumn{3}{c}{$W^- n/(n-1)$}  & \multicolumn{3}{c}{$Z n/(n-1)$} \\
        \noalign{\vskip 2mm}
        \colrule
        \noalign{\vskip 2mm}
        $n$ & LO & NLO & \MINLOp{} & LO & NLO & \MINLOp{}
        & LO & NLO & \MINLOp{} \\
        \noalign{\vskip 2mm}
        \colrule
        \noalign{\vskip 2mm}
        2 &  $0.3351(5)$ & $0.259(1)$ & $0.265(1)$ &$0.3172(4)$ & $0.248(1)$ &$0.253(1)$& $0.3219(4) $ & $0.2578(5)$ &$0.2624(5)$ \\
        3 &  $0.2894(6)$ & $0.250(2)$ &$0.247(2)$ &$0.2755(5)$ & $0.238(1)$ &$0.235(2)$& $0.2901(4)$ & $0.249(1)$&$0.247(1)$\\
        4 &  $0.288(1)$ & $0.245(5)$ &$0.244(5)$ &$0.2694(7)$ & $0.239(4)$ &$0.244(5)$& $0.2823(4)$ & $0.254(3)$&$0.254(4)$\\
        5 &  $0.279(3)$ & $0.252(12)$ &$0.251(11)$ &  $0.261(1)$ & $0.240(8)$ &$0.238(8)$& ---
        & --- & ---   
      \end{tabular}
    \end{ruledtabular}
  \end{center}
\caption{LO, NLO and \MINLOp{} QCD jet production ratios for $W^{\pm}$ as well as
$Z/\gamma^*$. The ratio is taken for a given process to that with one fewer jet.
The setup is specified in section~\ref{sec_kin}. The
number in parenthesis next to the ratio gives the corresponding statistical
integration error.\label{tab_jet_prod_total_xs} }
\end{table}

We observe that LO cross sections for the scale choice $\mu_0=\HTpartonicp/2$ have a monotonic increase of scale dependence
from about 20\% for $V+2$ jet up to 50\% for $V+5$ jets. We notice that the
LO scale dependence for $V+1$ jet, which appears at around 4\% is not
representative of the associated theoretical uncertainties, in particular due to
kinematical constraints that are released at NLO. That is, at LO the $p_{\rm T}$
of the vector boson matches the one from the unique jet. Real contributions at
NLO release this constraint, producing a soft enhancement that tends to produce
large corrections~\cite{Bauer:2009km,W3jDistributions,RubinSalamSapeta}.

The central scale choice for the dynamical scale $\mu_0=\HTpartonicp/2$
falls near the plateau of the NLO scale dependence. This makes the
uncertainty estimates based on lower/upper values of the cross sections seem slightly
small. If we quote the absolute deviations with respect to this value, then we
can estimate NLO scale sensitivity at the order of 10\% (running from about 6\%
to 16\%, depending on multiplicity). 

We find similar total cross sections with \MILOp{} to the corresponding LO
results. Nevertheless,
the absolute predictions are larger for \MILOp{}, and the excess increases slightly
with multiplicity. In the case of \MINLOp{} results, the central predictions agree
well with the NLO results. Moreover, their ratio is rather
stable as a function of the jet multiplicity.

In figures~\ref{fig_Wp_all_scales}-\ref{fig_Z_all_scales} we display total
cross sections and scale variations for all scales considered in this analysis.
It can be seen that in
particular the \MILOp{} and \MINLOp{} predictions exhibit a considerable variation,
both in their central value and in the associated conventional scale uncertainty.
This can be traced back to the procedure for the identification of ordered clustering
hierarchies, and the associated value of the scale of the core interaction, $\mu_{\rm core}$
(cf. Sec.~\ref{sec_minlo_gen}). Since we require the branching history to be ordered,
we must terminate the clustering as soon as an inverted scale hierarchy is encountered.
This biases the scale choice for events with many hard scales towards $\mu_{\rm core}$,
and therefore the precise definition of $\mu_{\rm core}$ plays a significant role~\cite{CKKWincl}.
The choice of $\HTpartonicpp$ increases $\mu_{\rm core}$ on average, thus permitting
more clusterings in high-multiplicity final states, and therefore inducing more associated
Sudakov form factors. On average this reduces both the central value of the prediction
and the related scale uncertainty.

Jet-production ratios have been shown to be a good handle for precision tests of
QCD. In these ratios many uncertainties associated with scale sensitivity and
PDF dependence cancel to a large extent. Also, in experimental analyses, systematic
uncertainties associated with luminosity measurements are canceled.  In
table~\ref{tab_jet_prod_total_xs} we observe a remarkable stability of these
ratios at NLO for both scale choices $\HTpartonicp/2$ and \MINLOp{}, all of them falling around a value of $0.25$.  This universality
is present for NLO results in $V+$jet even though the corresponding LO results
deviate considerably. The universality of the jet production ratios can be very
helpful for tests of the SM and can even be exploited to make extrapolations of
total and differential cross sections to large jet-multiplicity
processes~\cite{Wratios}.

\begin{figure}[t]
\begin{center}
\includegraphics[clip,scale=0.6]{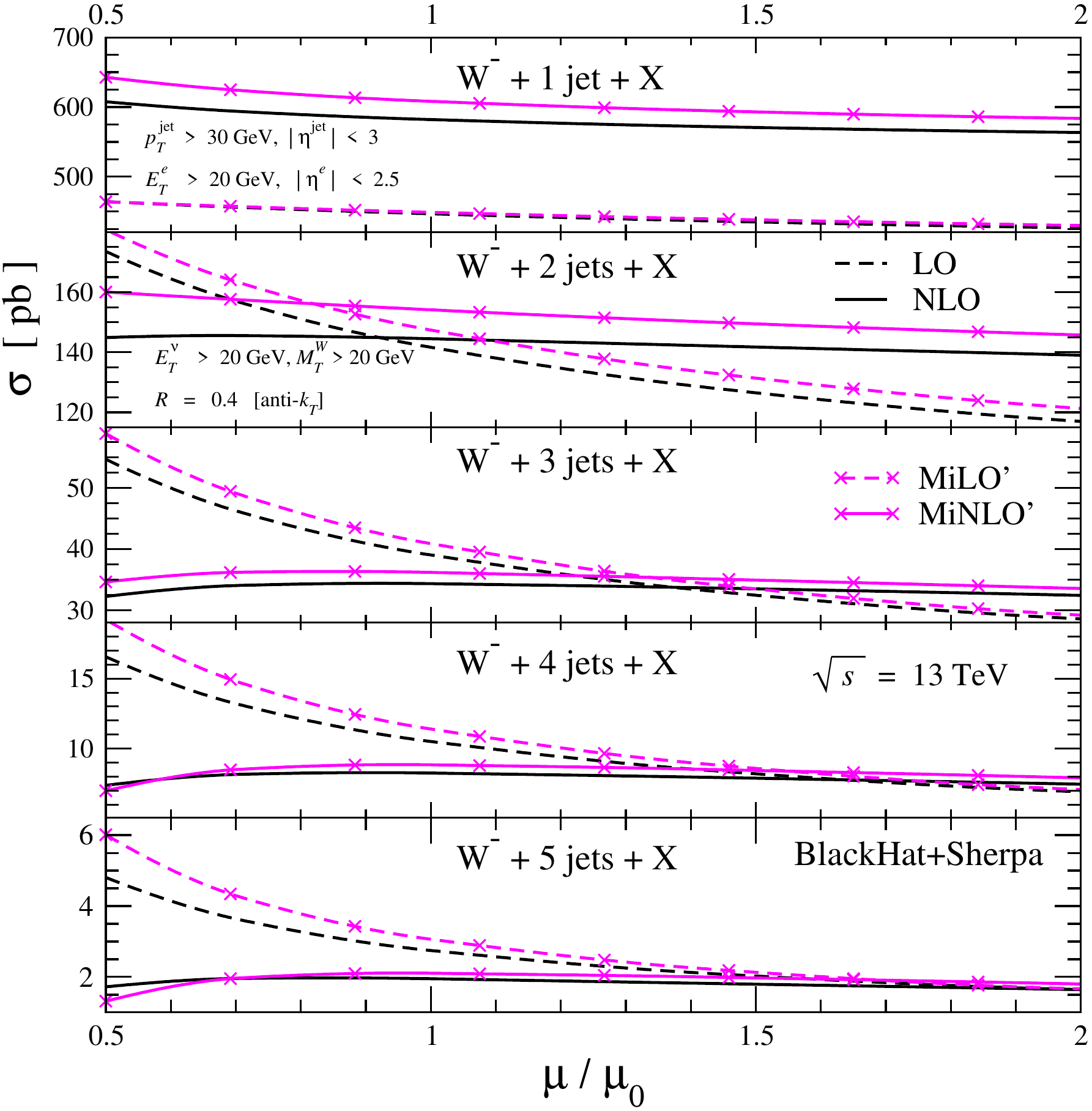}
\end{center}
\caption{The renormalization- and factorization-scale dependence of total cross
  sections for inclusive $W^-+1,2,3,4,5$-jet production. For each
  multiplicity, we show the dependence of predictions at LO as dashed (black) lines
  and at NLO as solid (black) lines with $\mu_0=\mu_{\rm r}=\mu_{\rm f}=\HTpartonicp/2$, while predictions for \MILOp{}
  are shown as dashed-crossed (magenta) lines and those for \MINLOp{} as solid-crossed (magenta) lines.}
\label{fig_Wmjets_sdep}
\end{figure}

\subsection{Scale Dependence}\label{wscale}
Figure~\ref{fig_Wmjets_sdep} displays the dependence of total cross sections
in $W^-$ production in association with up to 5 jets on the renormalization
and factorization scales. Results for the production of $W^+$ or $Z$ in
association with jets are similar. Although LO and \MILOp{} results are very sensitive
to these scales, we find a remarkable stability for NLO and \MINLOp{} results.
Note that the central prediction lies at the plateau of the NLO curves for
$W^-+$3,4 and 5 jets, thus minimizing the scale variations
(for a discussion of this effect, see~\cite{W3jDistributions}).

LO results obtained with the two scale choices appear as largely consistent,
concerning both normalization and scale sensitivity. \MILOp{} results are
slightly larger compared to LO results obtained with $\HTpartonicp/2$.
This indicates that the small average renormalization scales
for large jet multiplicity in \MILOp{} and the correspondingly large strong couplings
are not entirely compensated by the suppression from Sudakov form
factors. The LO results obtained with the two scale choices 
differ in their variation. At low multiplicity, the scale uncertainty
associated to LO and \MILOp{} results are comparable, but for increased
multiplicity the \MILOp{} results exhibit larger scale variations.

We observe that the NLO results obtained with both dynamical scales are mostly
consistent. For multiplicities with more than $2$ jets, the differences between
the two scale choices lie within the respective factor-two scale variations. In
the case of $W+2$-jet production, an approximately 15\% discrepancy appears,
which can be taken as an estimate for the total scale uncertainty
associated with the prediction. Furthermore, we observe that with increasing
multiplicity, the bands associated with the uncertainty of the respective scale
choice seem to behave differently between $\HTpartonicp/2$ and the \MINLOp{}
scheme. In general, the scale uncertainty obtained by factor-two variations
grows with multiplicity, e.g. for $\HTpartonicp/2$, from around $4.5\%$ for
$W^-+2j$ to about $15\%$ for $W^-+5j$ production. It is now interesting to
observe that this increase in scale uncertainty is more pronounced for the
\MINLOp{} scale choice, where the uncertainty grows from around $6.5\%$
($W^-+2j$) to around $35\%$ ($W^-+5j$).

\begin{figure}[t]

\includegraphics[clip,scale=0.98]{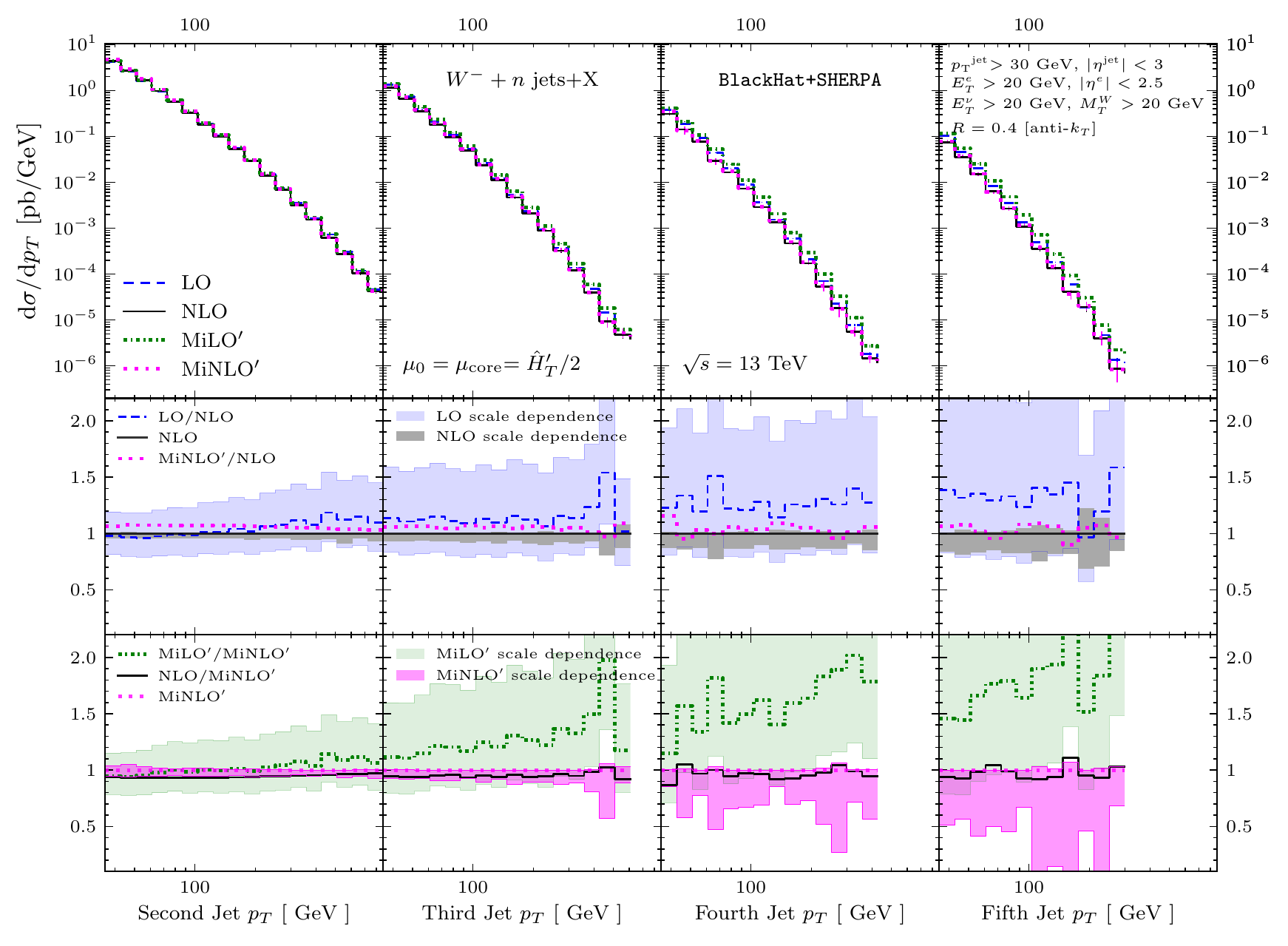}
\caption{The $\pT$ distribution of the softest jet in inclusive
  samples of $W^-+n$ jets ($n=2,3,4,5$) at the LHC at $\sqrt{s}=13$~TeV. In the upper panels, NLO
predictions are shown as solid (black) lines, \MINLOp{} predictions as dotted
(magenta) lines, while LO predictions are shown as dashed (blue) lines
and \MILOp{} predictions as dash-dotted (green) lines. The
central panels show the predictions for both the LO and \MINLOp{} distribution as
well as the scale-dependence bands normalized to the central NLO
prediction in dark gray for NLO and blue for LO. Similarly at the
bottom, we show predictions for \MILOp{} and NLO distributions as well
as scale dependence bands for
\MILOp{} in green and
\MINLOp{} in magenta normalized to the central \MINLOp{} predictions.
}
\label{fig_Wmnjpt}
\end{figure}

\begin{figure}[t]

\includegraphics[clip,scale=0.98]{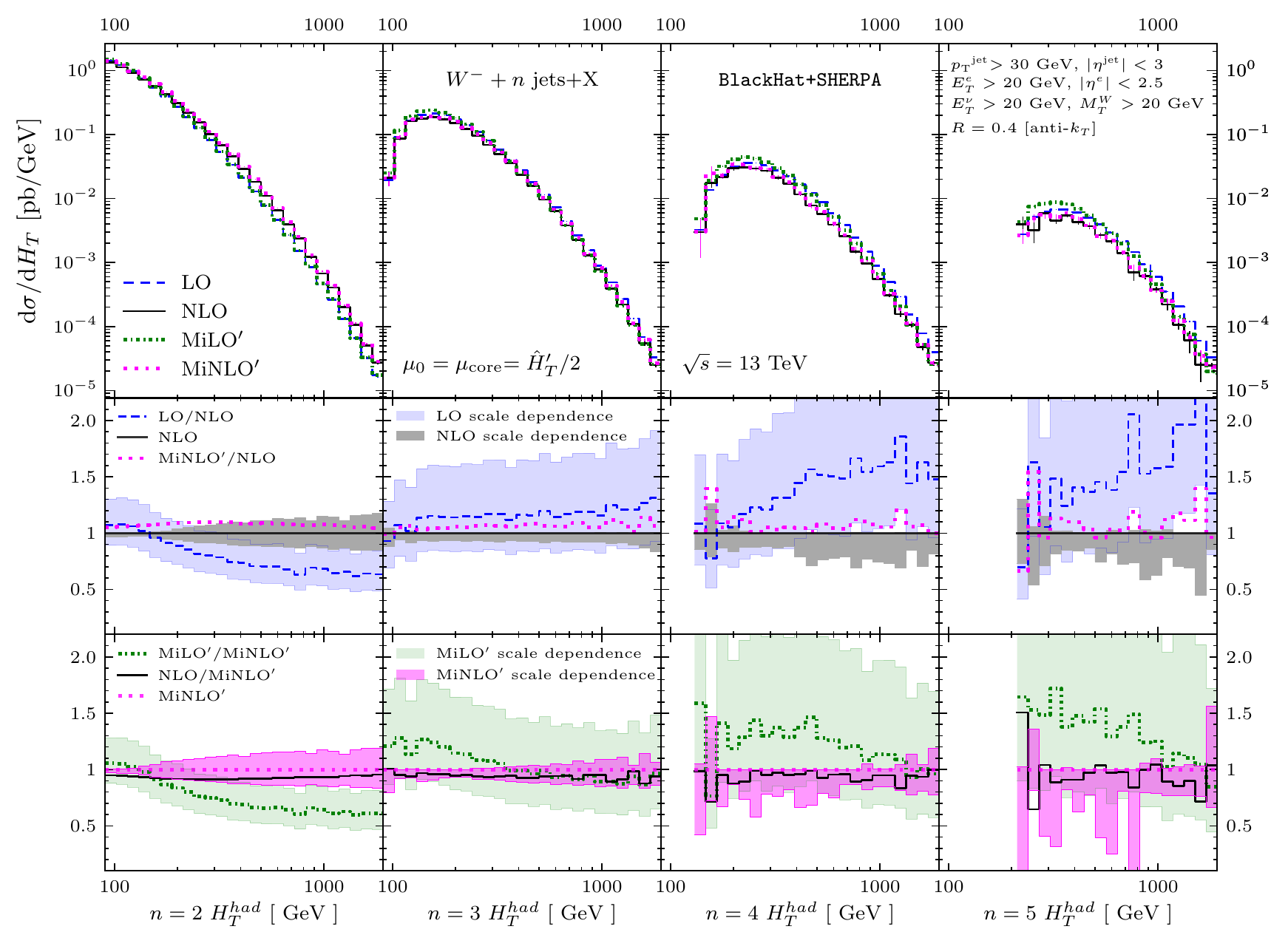}
\caption{Distribution in the total hadronic transverse energy $H_T$ in inclusive samples
of $W^-+2,3,4,5$-jets. Format as in Figure~\ref{fig_Wmnjpt}.
}
\label{fig_Wmjets_HT}
\end{figure}

\begin{figure}[t]

\includegraphics[clip,scale=1]{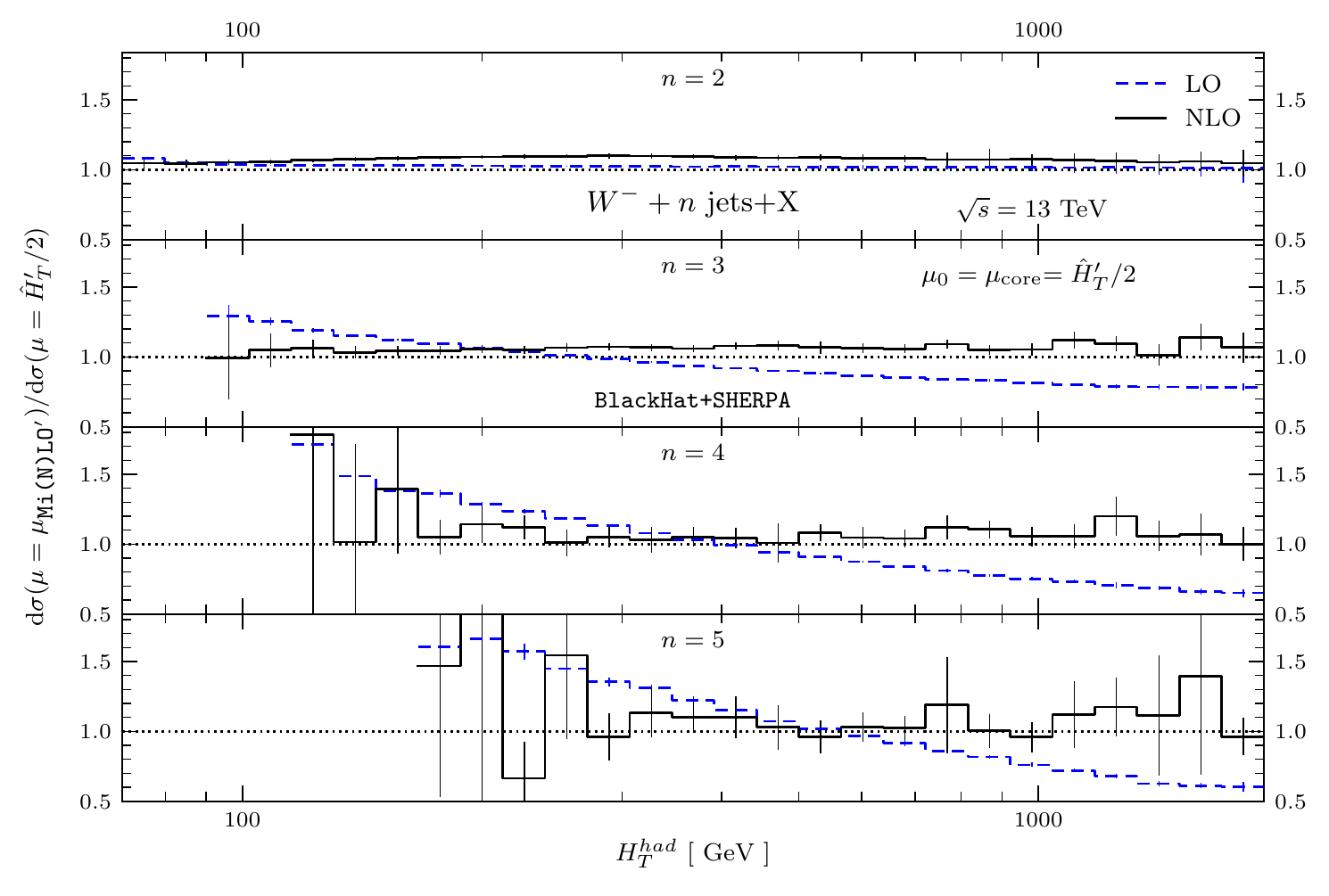}
\caption{\MINLOp{} to $\HTpartonicp/2$ ratio at both LO and NLO for
  the distribution in the total hadronic transverse energy $H_T$ in
  inclusive samples of \Wmjn-jet production
($n=2,3,4$ and $5$).
}
\label{fig_MINLO_HTp_ratios_HT}
\end{figure}

\subsection{Differential distributions}
\label{diffxsw}

In this section we analyze several differential distributions and the impact
that higher-order corrections have on fixed-order predictions over phase space. In
what follows we show results only for the $W^-$ weak vector boson,
as in general the structure of QCD corrections are very similar for the different vector bosons $W^\pm$ and
$Z$.
Also, we only include (N)LO and \MILNLOp{} results, as the associated results
with changing $\HTpartonic/2$ to $\HTpartonicpp$ are rather consistent.

Figure~\ref{fig_Wmnjpt} displays the $n$-th jet transverse momentum spectra in the
calculation of $W^-$+$n$ jets for $n=2$ to $5$. The solid (black) lines show NLO predictions, the dotted (magenta) lines
\MINLOp{} predictions, while the dashed (blue) lines show LO predictions
and the dash-dotted lines \MILOp{} predictions. The error bars represent the estimate of the statistical integration
errors. 
The middle panels show ratios to the NLO result including scale
dependence bands at LO and NLO. Similarly, the lower panels show
ratios to the \MINLOp{} results and scale dependences for \MILOp{} and \MINLOp{}.
Previous studies at lower energies (see for example~\cite{W5j}) have
shown that the $n$-th jet $p_{\rm T}$ spectrum in an inclusive $V+n$ jets sample
tends to have rather small distortions due to QCD corrections (as long as $n>1$).
Our current study confirms this result, with the LO to NLO ratios being flat over
a wide $p_T$ range. It is clear that although LO results employing an
$\HTpartonicp/2$ dynamical scale have similar shapes to the NLO results, their
normalization is very badly estimated, following the trends described in the
previous section on total cross sections. We also notice that the \MINLOp{}
results are in very good agreement with the NLO results in both shape and
normalization over the seven orders of magnitude shown for the differential
cross sections. This confirms that the predictions from NLO
results for these observables are under good theoretical control, with
uncertainties of order 15\% (considering the $\HTpartonicp/2$ scale-dependence
band as well as its deviation with respect to the \MINLOp{} result).

\begin{figure}[t]

\includegraphics[clip,scale=0.98]{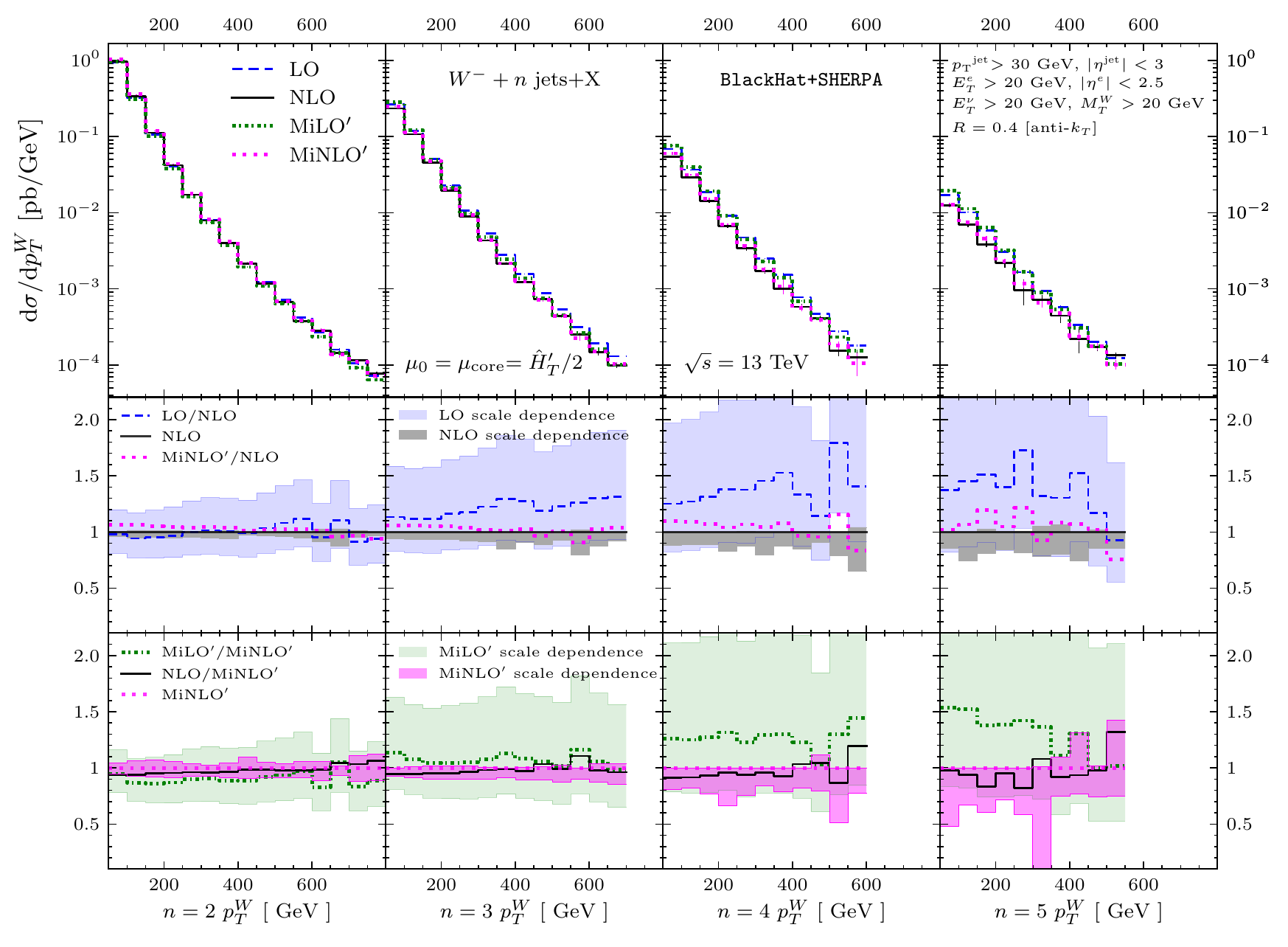}
\caption{Distribution in the transverse energy of the $W^-$ boson in inclusive samples
of $W^-+2,3,4,5$-jets. Format as in Figure~\ref{fig_Wmnjpt}.}
\label{fig_Wmjets_ptw}
\end{figure}
We study the distribution in the total jet hadronic transverse energy
$H_{\rm T}$ in Figure~\ref{fig_Wmjets_HT} as well as ratios thereof
between the \MINLOp{} and $\HTpartonicp/2$ scale choices in Figure~\ref{fig_MINLO_HTp_ratios_HT}. The format of Figure~\ref{fig_Wmjets_HT} is the same as
in Figure~\ref{fig_Wmnjpt}. We show in side-by-side panels results for $W^-+n$
jets, with $n=2$, $3$, $4$ and $5$. 
In general we find that for $n\ge 3$, the
corrections to the $H_{\rm T}$ spectrum change the shape of the
distribution only mildly, which can be seen by looking at both the LO to NLO ratios as well
as the \MINLOp{} to NLO ratios. Notice that the fluctuations at NLO for small
values of $H_{\rm T}$, as we increase multiplicity, are just due to the fact
that near threshold the integration errors grow large. On the other hand, the
$n=2$ LO predictions have a large shape difference compared to the NLO results. These
changes are similar to the corresponding observable for $n=1$ for which it is well
known that large corrections appear from configurations with many jets in the
final state~\cite{RubinSalamSapeta}. The widening of the NLO scale band indeed shows that real
contributions are large in the tail of the distribution, making this observable
sensitive to quantum corrections (an associated observation for the same
observable but at a hadron collider running at 100 TeV was made
in~\cite{FCC100TeV}). In principle a computation of NNLO QCD correction to $V+2$
jets would be desirable in order to stabilize the predictions for this highly relevant
process.

We notice a different behavior regarding the comparisons between
\MILOp{} and \MINLOp{} in the lower panels of Figure~\ref{fig_Wmjets_HT} as well
as in the \MILOp{} to LO ratios
shown in Figure~\ref{fig_MINLO_HTp_ratios_HT}. This arises from the fact that very
large values of $H_T$ are mostly generated in events of di-jet type with largely
disparate scales of jet production, i.e.\ $p_{T,j1}\approx p_{T,j1}\gg p_{T,j3},\ldots,p_{T,jn}$.
This induces Sudakov suppression factors that reduce the corresponding high-$H_T$
tails and improve the agreement with the NLO prediction. At NLO the
ratios in Figure~\ref{fig_MINLO_HTp_ratios_HT} are quite stable and around 1.

\begin{figure}[t]
  \includegraphics[clip,scale=0.82]{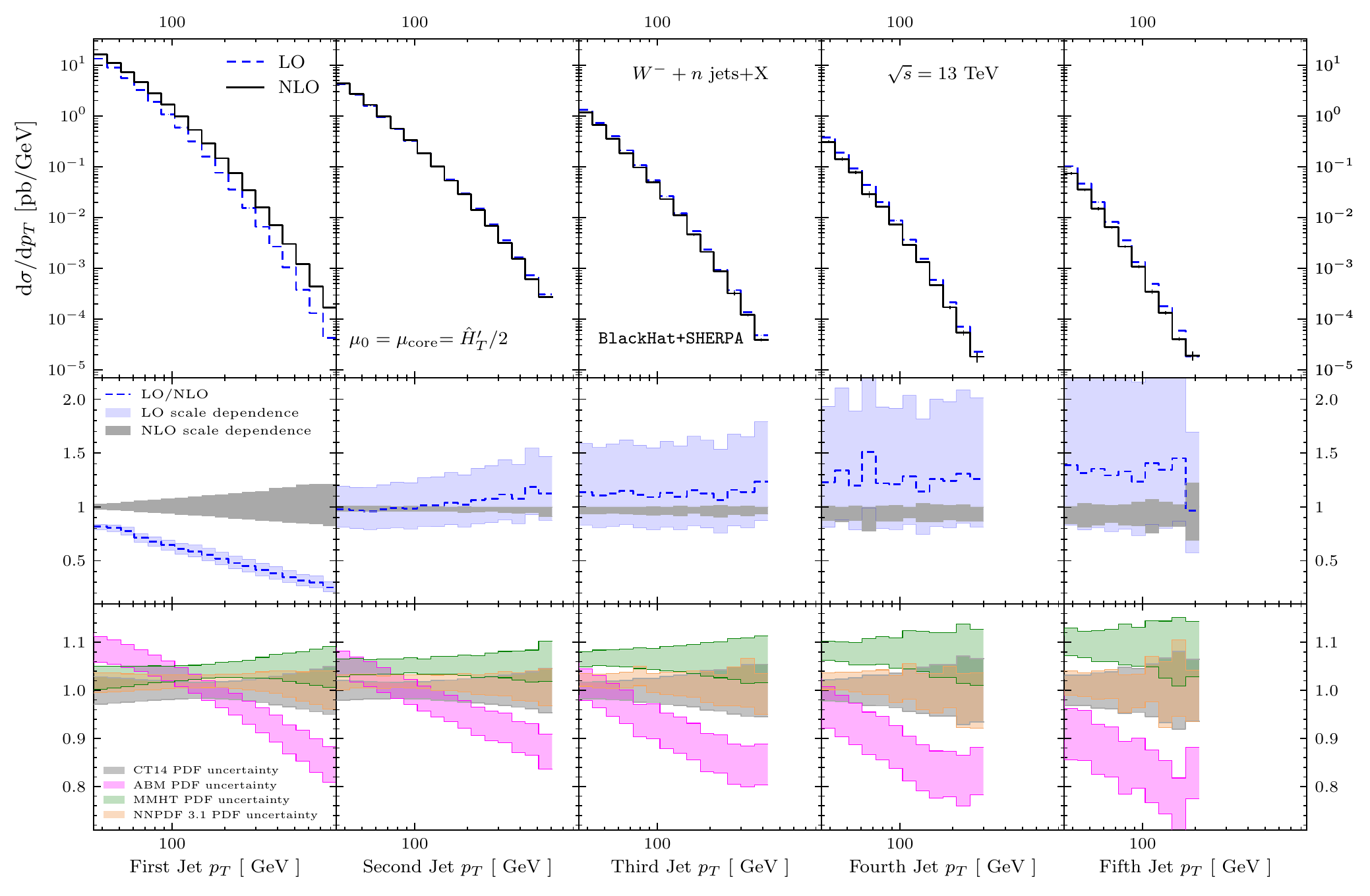}
\caption{PDF uncertainty for the $\pT$ distribution of the softest jet in an inclusive sample
of $W^-+n$ jets ($n=1,2,3,4,5$) at the LHC at $\sqrt{s}=13$~TeV. In the
  upper panels, the NLO predictions are shown as solid (black) lines,
  while the LO predictions are shown as dashed (blue) lines. The
central panels show the scale-dependence bands normalized to the
central NLO prediction in dark gray for NLO and blue for LO. The lower
panels show the NLO PDF uncertainty for the CT14 (gray), ABM
(magenta), MMHT (green) and NNPDF 3.1 (orange) PDF
sets, normalized to the results obtained with CT14.}
\label{fig_WmnjptPDF}
\end{figure}

Figure~\ref{fig_Wmjets_ptw} shows that the differential cross section of
the transverse momentum of the $W^-$ boson is well behaved under quantum
corrections. Indeed, these steeply falling distributions show very similar
shapes for LO, NLO, \MILOp{} and \MINLOp{} results. This distribution is interesting from
the experimental point of view, not only for its relation to several searches
for BSM physics, but also because it is associated to having under control
known missing energy signatures.

\begin{table}
\begin{center}
  \begin{ruledtabular}
    {\footnotesize
    \begin{tabular}{cccccc}
      \noalign{\vskip 1mm}
      nbr. of jets  &  1&	2&	3&	4&	5\\
      \noalign{\vskip 2mm}
      \colrule
      \noalign{\vskip 2mm}
      CT14      &$1.000$  &$1.000$  &$1.000$  &$1.000$   &$1.000$  \\
      ABM       &$1.070$ &$1.039$  &$1.000$ &$0.960$  &$0.920$ \\
      MMHT      &$1.029$ &$1.049$  &$1.066$ &$1.080$  &$1.095$ \\
      NNPDF 3.1 &$1.016$ &$1.018$  &$1.020$ &$1.020$  &$1.016$  \\
    \end{tabular}}
  \end{ruledtabular}
\end{center}
\caption{PDF variations of the NLO QCD total cross sections for inclusive
$W^-+1,2,3,4,5$-jet production normalized to the
results for CT14. This data was generated by creating a fastNLO
table \cite{Britzger:2012bs} from the \ntuple{} data.}
\label{tab_Wmj_pdf_central}
\end{table}

In Figure~\ref{fig_WmnjptPDF} we explore uncertainties associated to
the PDFs and show the PDF uncertainty bands for $n$th-jet $p_{\rm T}$ spectra
in an inclusive sample of $W^-+$jets for $\HTpartonicp/2$, analogous to the distributions shown in
Figure~\ref{fig_Wmnjpt}. The additional bottom panel shows the NLO PDF uncertainty
bands for the CT14 (gray)~\cite{CT14}, ABM (magenta)~\cite{ABM}, MMHT
(green)~\cite{MMHT} and
NNPDF 3.1 (orange)~\cite{NNPDF} PDF
sets, and we normalize to the central value obtained with CT14. 
The data for Figure~\ref{fig_WmnjptPDF} was generated by
creating a fastNLO table~\cite{Britzger:2012bs} from the \ntuple{} data. We also show the
central values for total cross sections obtained with the different
choices of PDF sets, normalized to the results for CT14, in table~\ref{tab_Wmj_pdf_central}.

We observe that PDF
uncertainties can reach values of up to 10\% and that most of the errors sets
overlap. Nevertheless,  both central value and uncertainty bands of the ABM
results lay outside the uncertainty bands of all other PDF sets. Also, the MMHT bands
lay systematically higher than the others, a trend that is more pronounced in the
large-multiplicity cases. All uncertainty bands increase at larger $p_{\rm T}$,
as the effective mass sampled for the corresponding events grows and the PDFs are evaluated for larger values of
the Bjorken $x$, with less data available to constrain the PDF
fits. PDF uncertainties are thus of the same order as NLO scale
uncertainties, in particular for the high-multiplicity processes,
where we observe a considerable spread between the different PDF sets. At the
level of normalized NLO QCD total cross sections this is shown in
table~\ref{tab_Wmj_pdf_central}.

\section{Cross Section Ratios}
\label{sec_ratios}
In this section we study a series of observable ratios, which feature reduced
uncertainties compared to basic observables such as cross sections or differential
distributions. In particular, experimental uncertainties related to jet energy
scale, lepton efficiency, acceptance and proton-proton luminosity should be
greatly reduced.  The observable ratios are also expected to suffer less from theoretical
uncertainties from uncalculated higher-order corrections. In consequence, these ratios
are helpful to better understand the structure of quantum corrections to
processes with a vector boson in association with multiple jets.  As shown
in~\cite{Wratios}, they exhibit certain universal features that
can be exploited in phenomenological studies at hadron colliders.  In this
section, we study the differential jet ratio for $W^-$ production as
well as differential flavor ratios of both $W^-/W^+$ and $Z/W$ production. 

\begin{figure}[t]

\includegraphics[clip,scale=0.75]{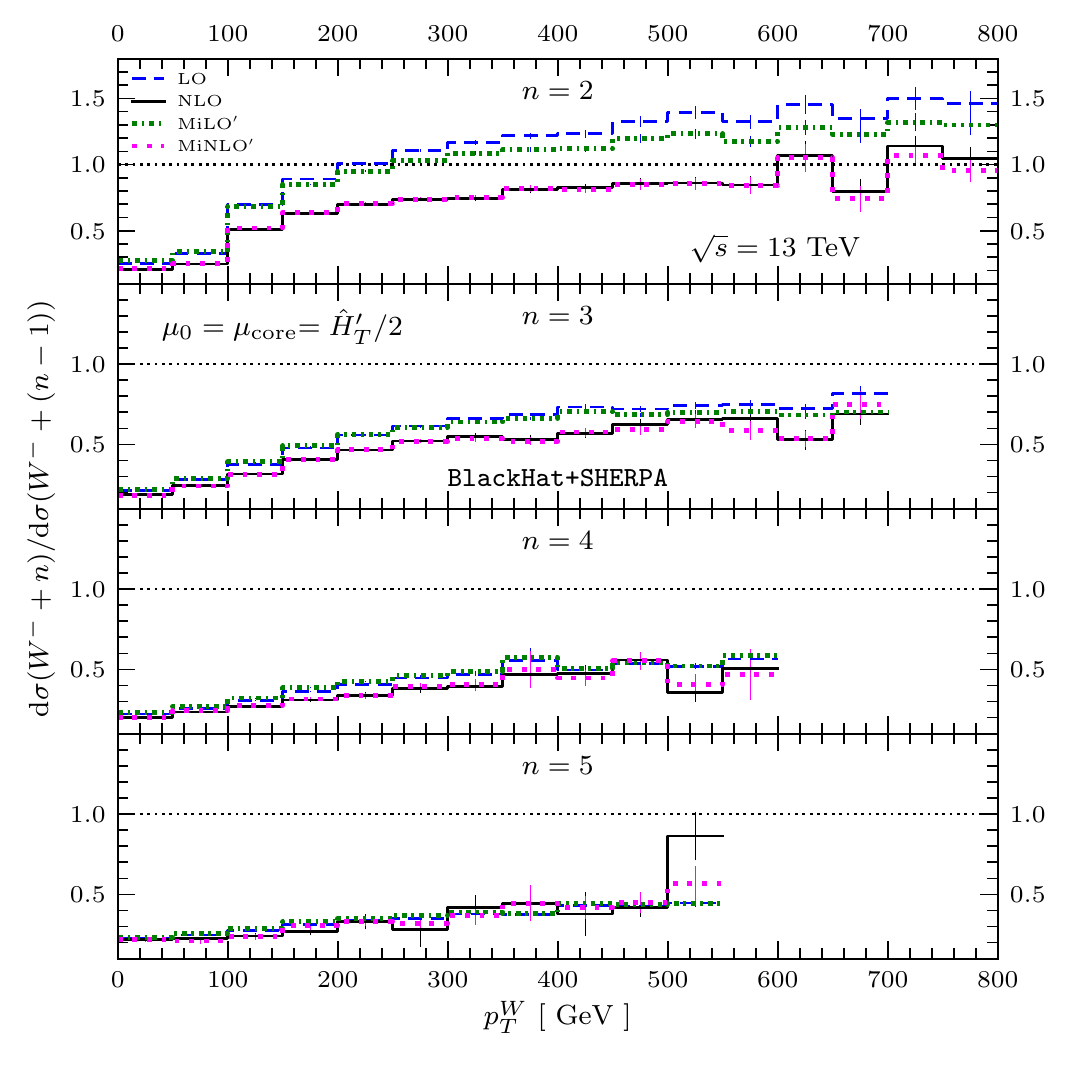}
\includegraphics[clip,scale=0.75]{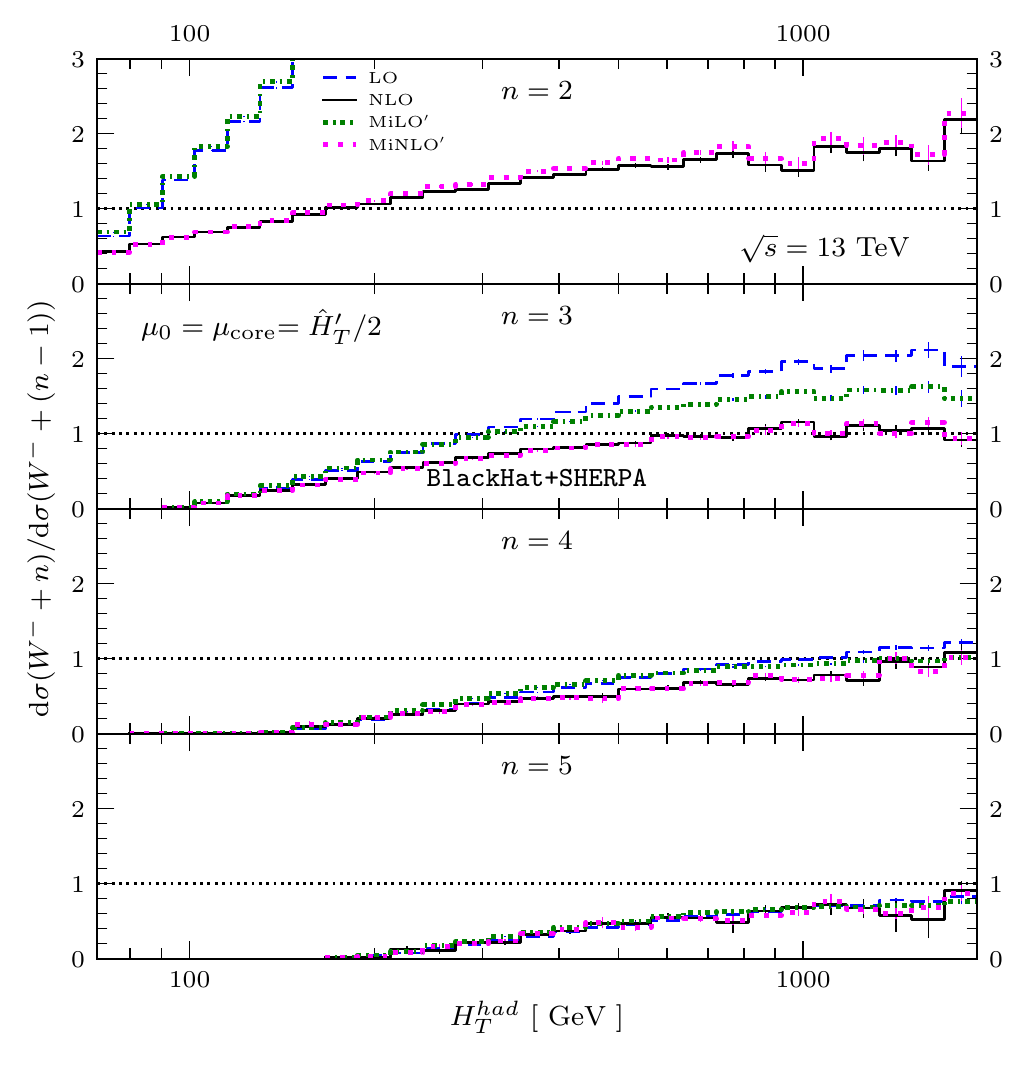}
\caption{The ratios of \Wmjn-jet to \Wmjnm-jet cross sections as a function of
the $W^-$ transverse momentum $p_{\rm T}^W$ on the left, and as a function of
the hadronic transverse energy $H_{\rm T}$ on the right. LO results are shown as dashed (blue) lines and NLO results as solid
(black) lines, while \MILOp{} results are shown as dash-dotted
(green) lines and \MINLOp{} results as dotted (magenta) lines. We present
results from $n=2$ (top panel) to $n=5$ (bottom panel).
}
\label{fig_jetrat}
\end{figure}

\subsection{Jet ratios}
\label{sec_jetratio}
We first examine the dependence of the jet-production ratio on the $W^-$-boson
transverse momentum $p_{\rm T}^W$ and total hadronic transverse energy $H_{\rm
T}$. We display the differential cross section ratios for the associated
production with up to 5 jets in Figure~\ref{fig_jetrat}, where the $p_{\rm T}^W$
distributions are shown in the left panel and the $H_{\rm T}$ distribution in
the right panel. The ratios are shown for both scale choices with
those for $\HTpartonicp/2$ shown at LO as dashed (blue)
lines and at NLO as solid (black) lines, while \MILOp{} ratios are
shown as dash-dotted (green) lines and those for \MINLOp{} as
dotted (magenta) lines.

As a function of $p_{\rm T}^W$, we observe that the \Wmjj-jet / \Wmj-jet ratio
($n=2$) shows large NLO corrections. This is mainly due to the large corrections
appearing at NLO~\cite{RubinSalamSapeta}, which are stabilized at
NNLO~\cite{VjetNNLO}.  In the lowest $p_{\rm T}$ region (up to the order of the
$W$ mass), the ratios lie around a value of $0.25$, roughly independent of the
number of jets. The NLO corrections are modest for all displayed multiplicities
in this region, being in agreement with the total cross sections displayed in
table~\ref{tab_Wmj_total_xs}. 
With higher $p_{\rm T}^W$, the ratios grow monotonically and for $n\geq 3$, they
stabilize for large values of $p_{\rm T}^W$.  For more jets, the increase is
less pronounced and the ratio stabilizes for $n=5$ around a value
$0.5$. For the low-multiplicity cases ($n\leq 3$), and in particular
for large values of $p_{\rm T}^W$, the ratio for
\MILOp{} lies below that for LO. The ratios for higher multiplicities as well
as all those of NLO and \MINLOp{} agree very well. 

Similar to the $p_{\rm T}^W$ ratios, we observe that the differential ratio in
$H_{\rm T}$ for $n=2$ is not stable with the fixed-order quantum
corrections. Even at NLO, the differential jet production ratio is
larger than 1 for a
large part of $H_{\rm T}$ shown. We would expect this
observable to stabilize as higher-multiplicity results are included either
through NNLO or higher-order calculations, or through multi-jet merging at NLO.
Around the threshold, all ratios lie at a value of the same order.
With increasing $H_{\rm T}$ they again
increase monotonically and at NLO stabilize for large values of $H_{\rm T}$.
Looking at the high $H_{\rm T}$ region, the ratios show a characteristic
behavior. Such events tend to be populated by multiple jets and the $H_{\rm T}$
distributions in these jet bins tend to overlap~\cite{Wratios}, which results in
the ratio tending to $1.0$. As for the previous distributions,
\MINLOp{} and NLO ratios agree very well, with a slight discrepancy
between LO and \MILOp{} ratios for the lower multiplicity cases.

\subsection{Flavor ratios}
\label{sec_ratiowmwp}
In this subsection we show flavor ratios of $W^-$ to $W^+$ production as well as
$Z$ to $W^+$ production. These ratios are of particular benefit as they are
reliable observables for the extraction of valence-quark PDF information for large values
of Bjorken $x$~\cite{Wratios,KomStirling}.

\begin{figure}[t]

\includegraphics[clip,scale=0.75]{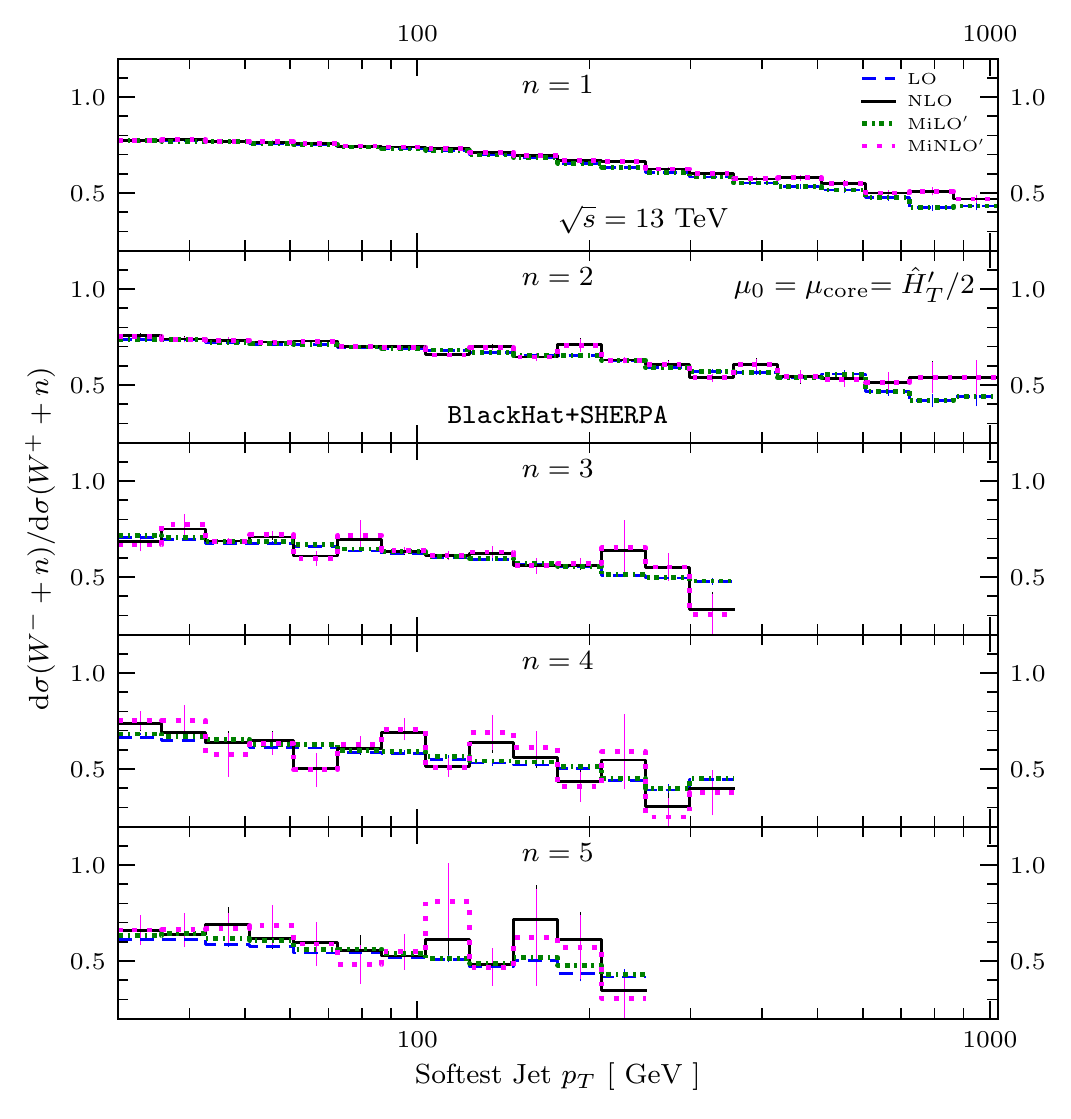}
\includegraphics[clip,scale=0.75]{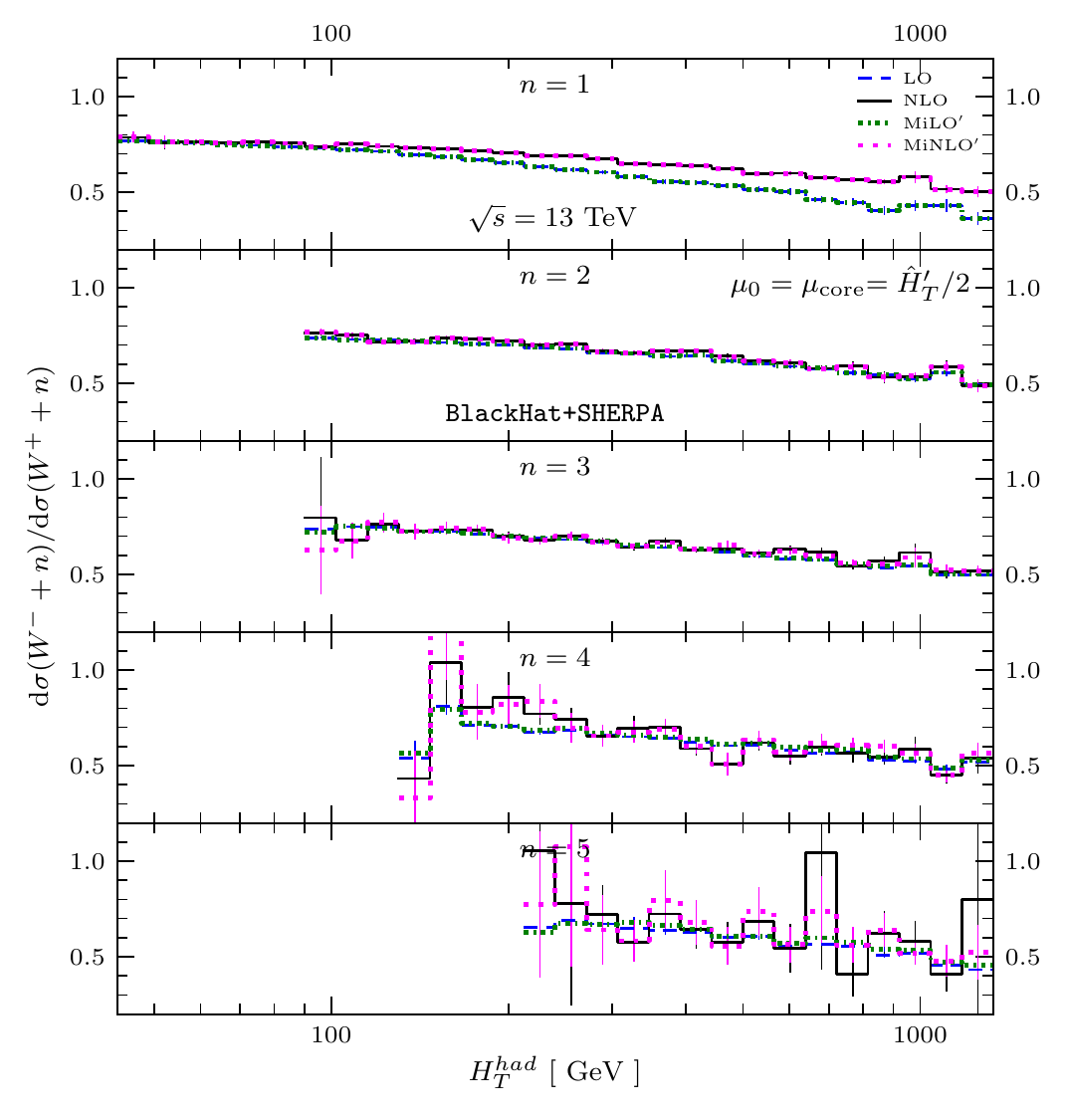}
\caption{The charge ratios of \Wmjn-jet to \Wpjn-jet cross sections as a function of
the softest-jet $p_{\rm T}$ on the left, and as a function of
the hadronic transverse energy $H_{\rm T}$ on the right, for results
from $n=2$ (top panel) to $n=5$ (bottom panel). Format as in
Figure~\ref{fig_jetrat}.
}
\label{fig_chargerat}
\end{figure}

\begin{figure}[t]

\includegraphics[clip,scale=0.73]{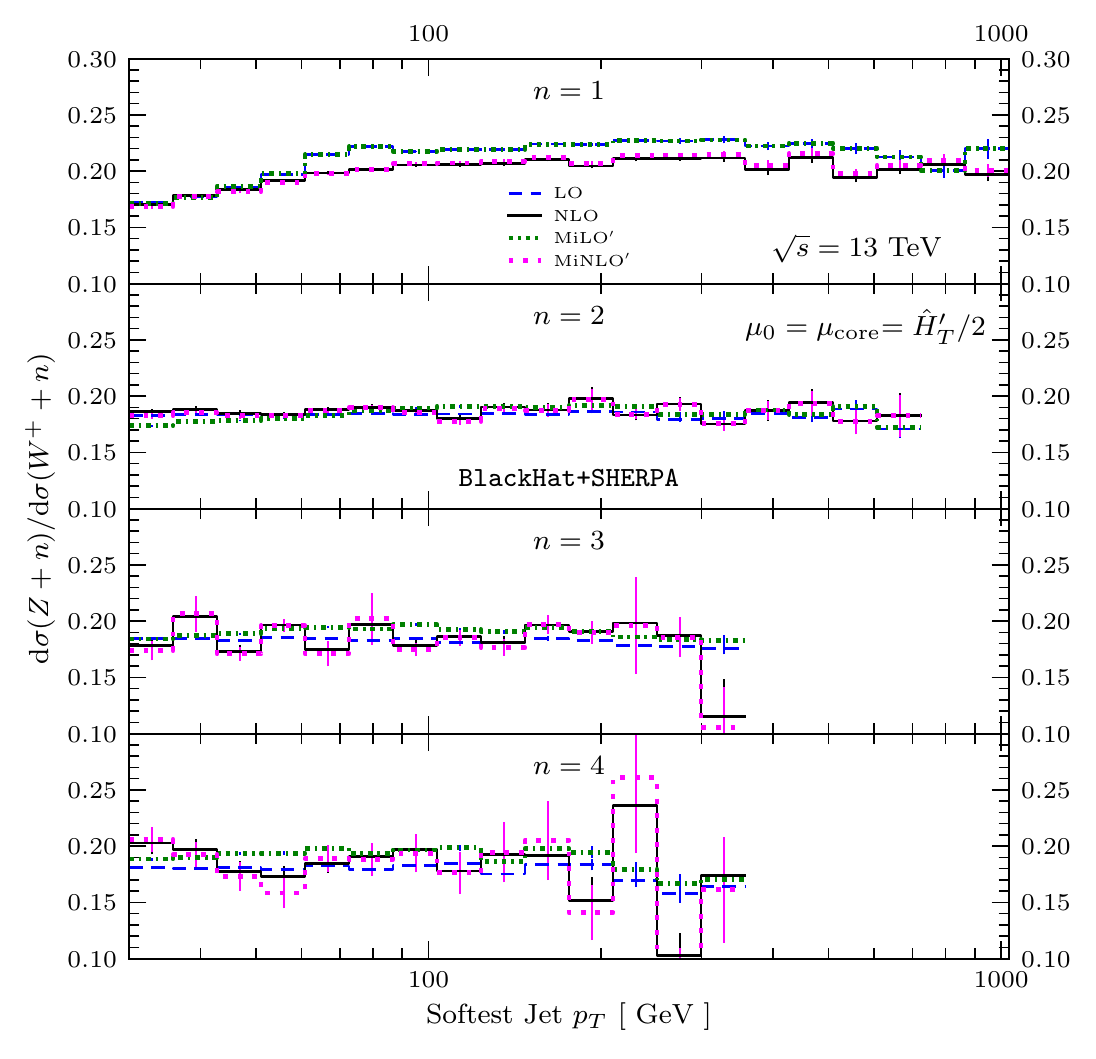}
\includegraphics[clip,scale=0.73]{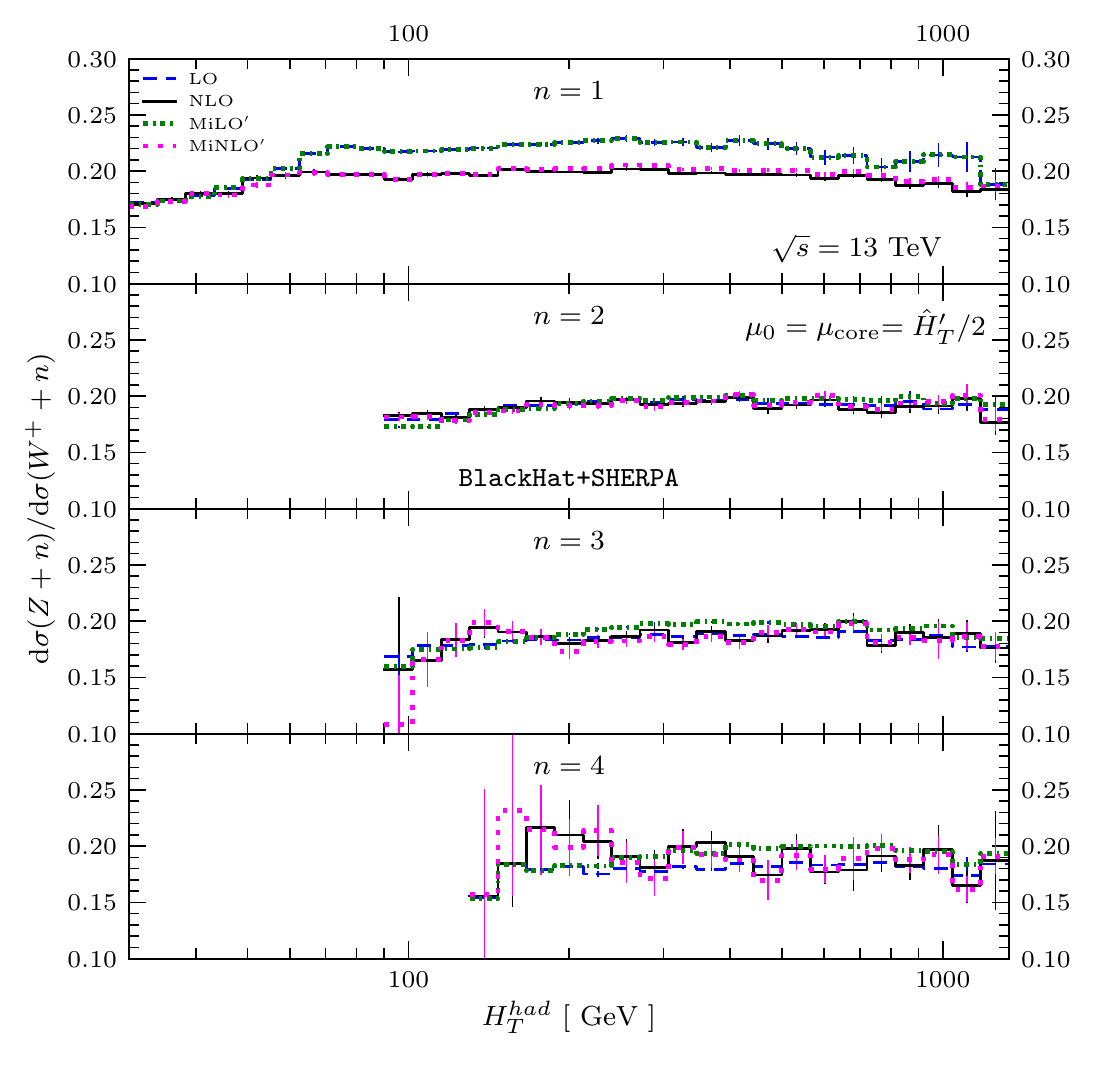}
\caption{The ratios of \Zjn-jet to \Wpjn-jet cross sections as a function of
the softest-jet $p_{\rm T}$ on the left, and as a function of
the hadronic transverse energy $H_{\rm T}$ on the right. We present
results from $n=1$ (top panel) to $n=4$ (bottom panel). Format as in
Figure~\ref{fig_jetrat}.
}
\label{fig_ZWp_rat}
\end{figure}

In Figures~\ref{fig_chargerat} and \ref{fig_ZWp_rat} we display differential ratios as a function of both, the transverse
momentum $p_{\rm T}$ of the softest jet (left panel) and of the total hadronic
energy $H_{\rm T}$ (right panel) for $W^-$ to $W^+$ accompanied with
up to five jets and $Z$ to $W^+$ with up to four jets
respectively. The ratios are shown for both scale choices with
those for $\HTpartonicp/2$ shown at LO as dashed (blue)
lines and at NLO as solid (black) lines, while \MILOp{} ratios are
shown as dash-dotted (green) lines and those for \MINLOp{} as
dotted (magenta) lines. Experimental
studies on the latter ratios have been made for example
in~\cite{ExptJetProductionRatioLHC}.

The $W^-/W^+$ ratios as a function of the transverse momentum for low values of $p_{\rm T}$ lie around a value of roughly the same
order, at about $0.8$ for $n=2$, for all
shown multiplicities $n$. With increasing $p_{\rm T}$, we see a monotonic
decrease of the corresponding ratio. We attribute this decrease to the
dominance of $u$ quarks over $d$ quarks at large values of Bjorken
$x$. The NLO corrections to the transverse momentum ratios are
mild and the results for both scale choices agree very well.

We observe a similar behavior for the differential ratios as a function of the total
hadronic energy $H_{\rm T}$. In particular, the ratios decrease
monotonically with increasing $H_{\rm T}$, 
and take values of the same order at low $H_{\rm T}$ for all
multiplicity, also the results for both scale choices agree very
well. However, we observe noticable NLO correction for the $n=1$ case in
the high $H_{\rm T}$ tail

The ratios of $Z$ to $W^+$ production in Figure~\ref{fig_ZWp_rat} are
quite flat over the full range of variables shown, which is in
contrast to the associated charge-asymmetry ratios in
Figure~\ref{fig_chargerat}. The values of $Z/W^+$ ratios as a function
of both the variables shown lie around
$0.2$ for all multiplicities $n$ shown. A small decrease of the $Z/W^+$ ratio can be observed in the low
$p_{\rm T}$ region in the ratio for $n=1$. Also for all the $Z/W^+$
ratios, the results for both scale choices agree very well. Our
results show that the quantum corrections in the $Z/W^+$ observable ratios are
quite mild, which makes them excellent choices for searches for new physics.

\section{Conclusions}
\label{ConclusionSection}

In this paper we presented NLO QCD predictions for the production of an electroweak
gauge boson with up to five jets for $W^\pm$, and with up to four jets for $Z$,
in the final state for the LHC with $\sqrt{s} = 13$ TeV, extending
previous predictions \cite{W5j,Z4j} to the energy of LHC Run II. 
Since these processes constitute an important background to many new physics
searches involving missing energy, as well as precise top-quark measurements,
the theoretical predictions we provide are an important ingredient to fully exploit
the discovery potential of the LHC in the upcoming years.

We observed that the scale dependence in NLO predictions for both total cross
sections and differential distributions is strongly reduced compared
to those at LO. Similar to the
$\sqrt{s} = 7$ TeV case, the scale dependence of total cross sections shrinks
from up to $50\%$ at LO to a sensitivity of about $15\%$ at NLO. 
We also compared results obtained with several other functional forms
of dynamical scales, most notably comparing results with fixed-order scales like
$\HTpartonicp/2$ with the \MINLOp{} (original formulation in \cite{MINLO}) reweighting procedure. This is the first
time this comparison is carried out in the context of processes with high jet
multiplicity. For total cross
sections, as well as for differential distributions, NLO results obtained with
$\HTpartonicp/2$ and \MINLOp{} are largely consistent in both shapes,
normalization and scale sensitivity.
In general the good agreement between both scale choices confirms that NLO
QCD predictions in high multiplicity processes give the first numerically reliable
predictions.

We furthermore computed several observable ratios and found that jet production ratios have an increased stability compared to the
results at $\sqrt{s} = 7$ TeV. We thereby have found a setup for which there is 
independence of the jet-production ratios on the number of jets.

The present study provides insight into the phenomenologically relevant process
class of electroweak gauge boson ($W^\pm,Z$) production plus up to five jets at
$\sqrt{s} = 13$ TeV. We look forward to more comparisons of our results, as the
ones recently shown in~\cite{AtlasVjets13}, with LHC data at the energy
frontier.

\section*{Acknowledgments}
We thank Z.~Bern, L.J.~Dixon, H.~Ita and D.A.~Kosower for many discussions.
We also thank Keith Hamilton for help with the detailed cross-check of our \MINLO{}
implementation, and for advice regarding the systematic uncertainties.
The work of F.R.A. and F.F.C. is supported by the Alexander von Humboldt
Foundation, in the framework of the Sofja Kovalevskaja Award 2014, endowed by
the German Federal Ministry of Education and Research.
S.H.'s work was supported by the US Department of Energy under contract
DE--AC02--76SF00515.
This work was performed on the bwUniCluster funded by the Ministry of Science,
Research and the Arts Baden-W\"urttemberg and the Universities of the State of
Baden-W\"urttemberg, Germany, within the framework program bwHP.
Parts of the calculation have been completed at the National Energy Research
Scientific Computing Center, a DOE Office of Science User Facility supported by
the Office of Science of the U.S. Department of Energy under Contract No.
DE--AC02--05CH11231.

\FloatBarrier


\end{document}